\documentclass[letterpaper,english,aps,manuscript]{revtex4}
\usepackage[T1]{fontenc}
\usepackage[latin1]{inputenc}
\usepackage{floatflt}
\usepackage{graphicx}
\usepackage{amssymb}

\makeatletter

\newcommand{\lyxdot}{.}

\usepackage{babel}
\makeatother
\begin{document}

\title{Effective potentials and electrostatic interactions in self-assembled
molecular bilayers II: the case of biological membranes.}

\author{Z. Gamba}

\affiliation{Department of Physics - CAC, Comisión Nacional de Energía Atómica,
Av. Libertador 8250, (1429) Buenos Aires, Argentina. }

\email{gamba@tandar.cnea.gov.ar. URL: http://www.tandar.cnea.gov.ar/~gamba}

\date{\today}

\begin{abstract}
We propose a very simple but realistic enough model which allows to
include a large number of molecules in molecular dynamics MD simulations
of these bilayers, but nevertheless taking into account molecular
charge distributions, flexible amphiphilic molecules and a reliable
model of water. All these parameters are essential in a nanoscopic
scale study of intermolecular and long range electrostatic interactions.
This model was previously used by us to simulate a Newton black film
and in this paper we extend our investigation to bilayers of the biological
membrane type. The electrostatic interactions are calculated using
Ewald sums and, for the macroscopic long range electrostatic interactions,
we use our previously proposed coarsed fit of the (perpendicular to
the bilayer plane) molecular charge distributions with gaussian distributions.
To study an unique biological membrane (not an stack of bilayers),
we propose a simple effective external potential that takes into account
the microscopic pair distribution functions of water and is used to
simulate the interaction with the surrounding water. The method of
effective macroscopic and external potentials is extremely simple
to implement in numerical simulations, and the spatial and temporal
charge inhomogeneities are then roughly taken into account. Molecular
dynamics simulations of several models of a single biological membrane,
of neutral or charged polar amphiphilics,  with or without water (using
the TIP5P intermolecular potential for water) are included.\\

\end{abstract}
\maketitle

\section{Introduction}

Amphiphilic bilayers play a key role in numerous problems of interest
in chemical physics, nano- and biotechnology. An amphiphilic molecule
consists of a non-polar hydrophobic flexible chain of the $(CH_{2})_{n}$
type, the 'tail', plus a hydrophilic section, a strongly polar 'head'
group. In aqueous solutions, the 'head' interacts with water and shields
the hydrophobic tails, so the the amphiphilics tend to nucleate in
miscelles or bilayers, depending on concentration and the 'head' group
size \cite{chandler}. A simple model of a biological membrane consists
in a bilayer of amphiphilic molecules, with their polar heads oriented
to the outside of the bilayer and strongly interacting with the surrounding
water. The opposite model of bilayer, with the water in the middle
and head groups pointing to the inside, also exists in nature and
are the soap bubbles films, or Newton black (NB) films, as they are
called in their thinnest states. 

We propose a very simple but 'realistic' enough model of amphiphilic
bilayers, so a large number of molecules can be included in the numerical
simulations and at the same time molecular charge distributions, flexible
amphiphilic molecules and a reliable model of water can be taken into
account. All these parameters are essential in a nanoscopic scale
study. Such amphiphilic bilayer models will be useful to obtain reliable
information on the effect of the {}``external parameters'' (like
surface tension, external pressure and temperature) on physical properties
of the membrane, as well as to address problems like the diffusion
and/or nucleation of guest molecules of technological, pharmaceutical
or ambiental relevance within these bilayers, the main area we are
interested in. In Ref. \cite{zg-bub08} we proposed an amphiphilic
model with the desired characteristics and applied it to the study
of the macroscopic electric field and intermolecular interactions
in a bilayer of the NB film type. Here we extend our study to the
case of a \emph{single} biological membrane model.\\

Many molecular dynamics MD simulations of biological membranes using
detailed all atom models of amphiphilic molecules have been performed.
For example, in ref. \cite{mem-mike-all.atom1} a sample of 64 DMPC
(dipalmitoylphosphatidylcholine) molecules and 1645 water molecules
were simulated for 2 nsec, using an interaction model that includes
Lennard-Jonnes LJ potentials and a set of charges distributed at atomic
sites. These type of MD calculations are extremely useful to obtain
reliable information not only on the membranes themselves but also
on the behavior of guest molecules. Their main problem is that they
are extremely lengthy and, due to the periodic boundary conditions
along the perpendicular to the bilayer and the relatively small number
of water molecules included, in most cases the simulated sample is,
in fact, a stack of bilayers. 

In general, biological membranes are more lengthy to simulate than
NB films with the same number of amphiphilics, due to the much larger
number of water molecules per amphiphilic needed to include in order
to obtain full hydration of their polar heads. Also the detailed atomic
description of a NB film amphiphilic is usually more simple than a
lipid of a biological membrane. That is the motivation behind the
study of extremely simple models \cite{farago,mem-potextra1,mem-mike-coarse.grain,mem-mike-allatom-coarsegrain,mempolym-mike-coarse-grain,mem-marcus-coarse.grain}
that, although do not include electrostatic interactions, have been
useful to study mesoscopic problems like thermal undulations and nucleation
of pores. We can include in this category the {}``water free'' models,
calculated with a bilayer of three sites linear rigid \cite{farago}
or flexible \cite{mem-potextra1} molecules, the site-site interactions
include repulsive and atractive Lennard- Jones potential terms but
not charges. Also {}``coarse grained'' membrane models have been
proposed \cite{mem-mike-coarse.grain,mem-mike-allatom-coarsegrain,mempolym-mike-coarse-grain,mem-marcus-coarse.grain},
which allow MD simulations in the order of hundreds of nanoseconds
in time scale and microns in space scale; in this way, even more complicated
problems, like membrane fusion, self-assembly of lipids and diblock
copolymers, have been addressed. \\

A further problem to solve, when studying biological membranes, is
that of the periodic boundary conditions along the perpendicular to
the bilayer. The usual approach is to include the largest possible
number of water molecules in the MD sample and to apply 3D periodic
boundary conditions \cite{scott}. In ref. \cite{bicapas-pastor},
this method was improved by using periodic boundary conditions and
a variable box size along the perpendicular to the bilayer plane.
In this way it is possible to work at constant surface tension (given
by the surface density of amphiphilics in the membrane) and at constant
external normal pressure, applied perpendicular to the bilayer. Nevertheless,
even using this type of approach and due to the periodic boundary
conditions, the thickness of the water layer in the MD box is usually
around 20 $\textrm{Å}$\cite{bicapas-pastor}, and therefore the simulation
is more adequated for the study of stacks of membranes.

In Ref. \cite{zg-bub08} we discussed the problem of these quasi 2D
highly charged bilayers and performed, as an example, the simulation
of a NB film. We used the Ewald method for the electrostatic intereractions,
but applying 2D periodic boundary conditions in the plane of the bilayer
plus a large empty space (along the perpendicular to the plane) in
the simulation box. We also proposed a novel, simple and more acurate
macroscopic electrostatic field for model bilayers and applied it
to the case of NB films. Our macroscopic field model goes beyond that
given by the total dipole moment of the sample, which on time average
is zero for this type of symmetrical bilayers. We showed that by representing
the higher order moments with a superposition of gaussians the macroscopic
field can be \emph{analytically} integrated, and therefore its calculation
easily implemented in a MD simulation (even in simulations of non-symmetrical
bilayers) \cite{zg-bub08}. \\

At variance with a soap bubbles film, the calculation of a \emph{single}
biological membrane implies a number of practical problems, related
to the interaction of the amphiphilic bilayer with the surrounding
water and how to perform an accurate calculation of the far from negligible
electrostatic interactions. To analyze the rôle of the water molecules
in the dynamics and stability of these aggregates we need, on one
hand, to really include a number of water molecules in the model,
because they diffuse around the polar head groups and not all of them
are 'outside' the bilayer. On the other hand, we have an upper limit
for the total number of water molecules that can be included in the
numerical simulations. 

There are many approaches to address the problem of the long ranged
electrostatic potential and the forces on a solved molecule due to
the surrounding solvent, as well as those arising from the diffusion
and collective motions of a large number of polar and/or charged molecules
in solution. The solutions range from the inclusion of a few water
molecules inside a dielectric cavity (the reaction field approach)
to the inclusion of a large number of water molecules in a sample,
using periodic boundary conditions and the Ewald's method to calculate
the electrostatic interactions. 

The reaction field approach to solve the long ranged electrostatic
interactions consists in consider each charge within a dielectric
cavity that can hold a small number of solute and solvent molecules
(for example, ref. \cite{roux,chandra99}), long range interactions
are avoided in this way, but the solvent behavior is strongly dependent
on the size and form of the nanopore. Moreover, the main problem of
the dielectric cavities (besides being a macroscopic approximation)
is that they delimit a constant volume with very  few water molecules
inside. Large fluctuations in all measured properties are due to the
fluctuating number of charges inside the cavity. The reaction field
method can be improved by applying a switching function that smooths
the dielectric boundary of the cavity, reducing thus the measured
charge fluctuation \cite{reac-field3}, or by defining a realistic
shape of the solute-solvent boundary (i. e. given by interlocking
spheres centered on solute atoms \cite{reactionfield-02}). Nevertheless,
remains the fact that a cavity in a homogeneus dielectric is a continuous
macroscopic approximation, and therefore an oversimplification of
the solute interactions with water at distances of a few angstroms
\cite{levin02}, this non-homogeneity is non-negligible, at least,
up to second neighbour distances. 

Here we propose to take into account the interaction of the amphiphilics
with the surrounding water by using a variable size 'realistic' cavity
that takes into account the microscopic pair distribution function
of water and the external pressure (the normal component for amphiphilic
bilayers).\\

In this paper we extend our study on NB films and present a few and
simple molecular models for the simulation a \emph{single} biological
bilayer. Electrostatic interactions, using Ewald sums and our proposed
macroscopic field model are taken into account in both types of films
\cite{zg-bub08}. Molecular dynamics (MD) simulations of a pure sample
of water and a few solved amphiphilic molecules in water, using the
TIP5P water intermolecular potential model, in order to obtain the
needed pair correlation functions are included. In the following sections
we present the external effective field of the surrounding water for
a \emph{single} biological membrane simulation, the macroscopic electric
field of biological membranes with periodic boundary conditions in
two directions perpendicular to the bilayer plane, the molecular dynamics
simulations performed for several proposed bilayers models: with and
without solved ions, and with or without water, as well as different
properties measured in them.

\section{Bulk water}

We selected the classical rigid TIP5P \cite{jorgensen1,jorgensen2}
molecular model for water. It consists of one LJ site ($\varepsilon=0.67$
kJ/mol, $\sigma=$ 3.12 $\textrm{Å}$) localized at the O and 4 charges,
two charges q$_{\textrm{H}}=0.241e$ are localized at the H atoms
and two q$_{Lp}=$-q$_{\textrm{H}}$  at the lone pairs. TIP5P gives
good results for the calculated energies, diffusion coeficient and
density $\rho$ as a function of temperature, including the anomaly
of the density near 4C and 1atm \cite{jorgensen1}, the X-ray scattering
of liquid water \cite{parrinello-w}, etc. The only exception is the
O-O pair correlation function $g_{2}(r)$, for which the first neighbor
 is located at a slightly shorter distance than the experimental one
\cite{jorgensen1}. \\

In order to obtain the needed pair distribution functions of water,
we performed a classical microcanonical MD simulation of pure water
at the 298K experimental density (29.9 $A^{3}$ per molecule). The
electrostatic interactions are calculated using 3D Ewald sums and
periodic boundary conditions are applied. The cut-off radius is 14
$\textrm{Å}$ and correction terms to the energy and pressure, due
to this finite cut-off, are taken into account. The minimum image
convention is applied. The equations of motion of the rigid water
molecules are integrated using the velocity Verlet algorithm for the
atomic displacements and the Shake and Rattle algorithms for the constant
bond length constraints on each molecule. The temperature, in this
simulation, is maintained using the Nosé-Hoover chains method \cite{nose-chains}.
The time step is of 1 fs., the sample is thermalized for 20 ps. and
measured in the followings 100 ps. As the lone pair interaction sites
are not coincident with atomic sites, the algorithm employed to translate
the forces from massless to massive sites is that of ref. \cite{algor6}.
The final version of the MD program is similar to that used in refs.
\cite{cyclobutane,sulfur-jcp01,sulfur-jcp2003}.\\

\begin{figure}
\includegraphics[scale=0.6]{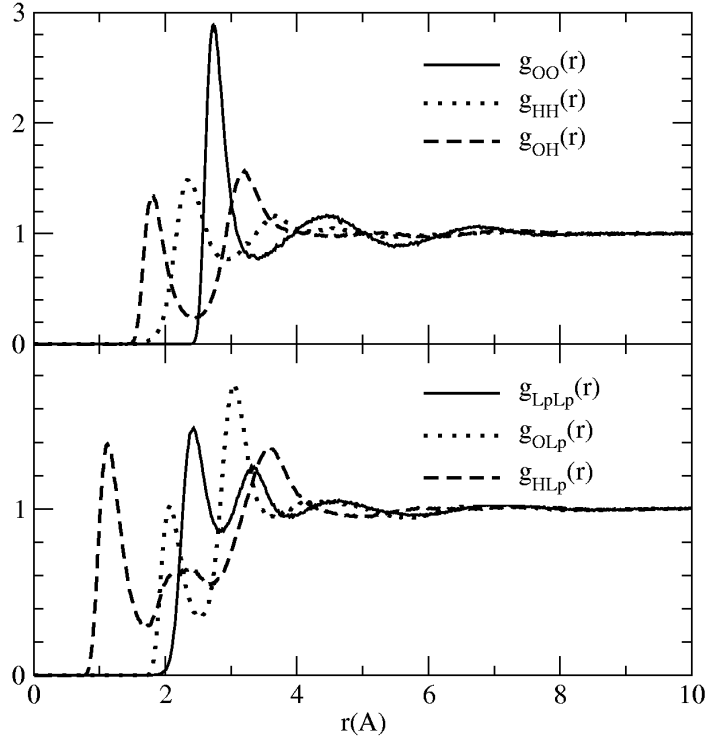}

\caption{The pair correlation functions for water at 298 K and 1 atm, calculated
with the TIP5P model.\label{cap:wsola-g2}}
\end{figure}

Fig. \ref{cap:wsola-g2} shows our calculated pair correlation functions,
obtained from two MD simulations of 864 and 1688 water molecules.
Note that we are including also the lone pair the pair correlation
functions with all other sites.

\begin{figure}
\includegraphics[bb=31bp 44bp 691bp 508bp,clip,scale=0.4]{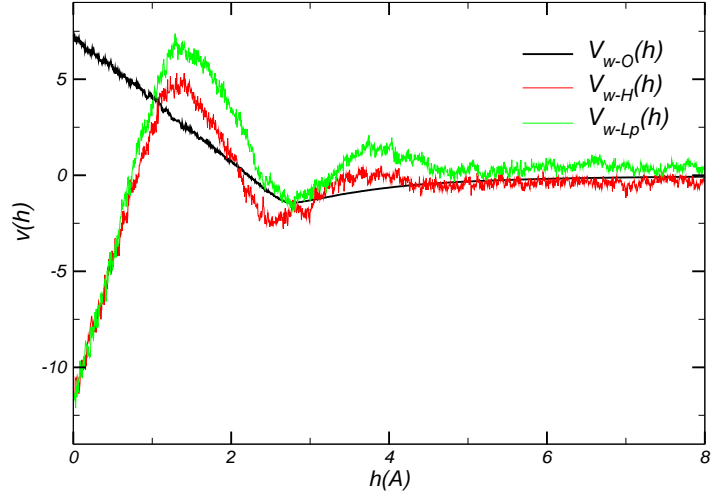}

\caption{Measured effective potential (Units: kJ/mol) of LJ and charged sites
with the 'liquid wall' (see text).\label{cap:MD-eff-pf}}
\end{figure}

In this MD simulation we also measured the interaction energy of a
molecule located at a distance \emph{$h_{z}$}>0 with all molecules
contained in the semi-volume defined by $z\leq0$. The number of water
molecules in half the MD box fluctuates as a function of time (with
a deviation of 2\%, at STP) and our measurement corresponds then to
the ($\rho$,P,T) ensemble. In Fig. \ref{cap:MD-eff-pf} we include
the measured effective potentials for LJ and charge - charge interactions
as a funcion of the distance $h_{z}$ between the site and the origin.
The Ewald constant was set equal to zero in this calculation. The
histograms for LJ atom-atom interactions converge to the values of
Fig. \ref{cap:MD-eff-pf} after a few ps., the electrostatic interactions
instead show very large fluctuations as a function of time, and the
averages are over 100 ps. The measured components of the electrostatic
forces along $x$ and $y$ axes show an averaged value of zero, with
a deviation of less than 5\% of the calculated maximum value for the
$z$ component. 

We found that the effective potential energies $U_{eff}(h)$ that
we measured in our MD run, can be reproduced by a mean field calculation
of the effective interaction potential of one molecule with a {}``wall''
at $z=0$, but taking into account the discrete distribution of particles
within the {}``semi-volume'' at $z<0$, as given by the corresponding
pair distribution function. That is, 

\begin{equation}
U_{eff}(h)=\int_{-\infty}^{0}dz\int_{-\infty}^{\infty}dx\int_{-\infty}^{\infty}dy\, U(r)\, g_{2}(r)\,,\:\, with\, r=\sqrt{(x^{2}+y^{2}+(z+h)^{2})}\label{eq:ueff}\end{equation}
 and the effective force is \emph{numerically} calculated:

\begin{equation}
F_{eff}(h)=-\frac{\partial}{\partial h}U_{eff}(h)\;.\label{eq:feff}\end{equation}
 Where \[
U(r)=4\varepsilon[(\frac{\sigma}{r})^{12}-(\frac{\sigma}{r})^{6}]\]
 for LJ interactions, and 

\[
U(r)=\frac{q_{i}q_{j}}{r}\]
 for charge $q_{i}$ - charge $q_{j}$ interactions.

Using the pair distribution functions of Fig. \ref{cap:wsola-g2},
the $U_{eff}(h)$ and $F_{eff}(h)$ functions were numerically integrated
for each pair of $site-site$ interaction. Note that, as we are dealing
with a disordered sample of neutral molecules, \emph{all} $g_{2}(r)$
pair correlation functions are approximately equal to 1 for $r\gtrsim10$A,
including those between charged sites (H and Lp sites of the TIP5P
model). These functions are used to calculate the effective interaction
of LJ sites and charges with the semi-volume at $z<0$. Therefore,
in a mean field approximation, both quantities $U_{eff}(h)$ and $F_{eff}(h)$
tend to zero for large distances $h$. 

\begin{figure}
\includegraphics[clip,scale=0.3]{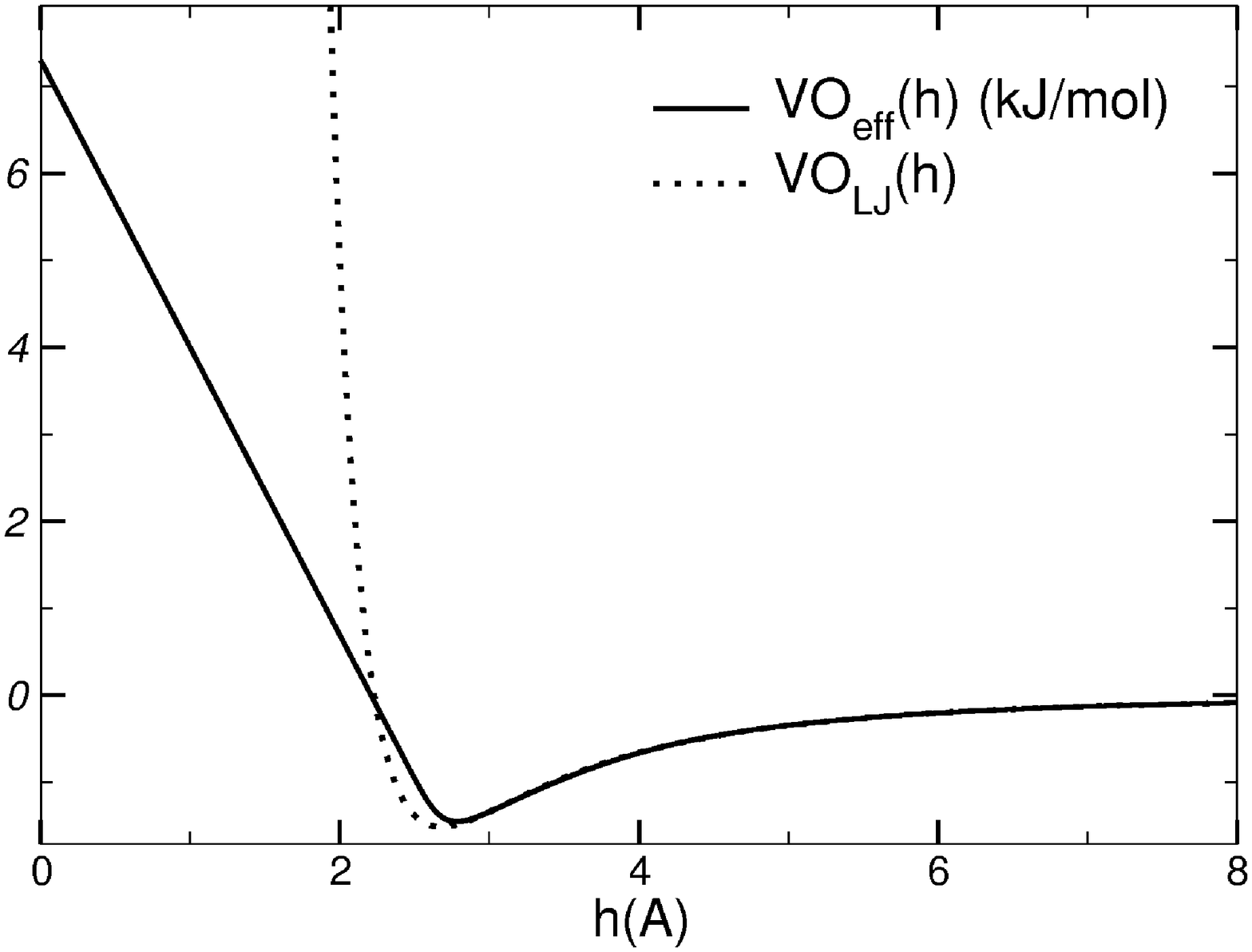}$\;\quad\:$\includegraphics[clip,scale=0.3]{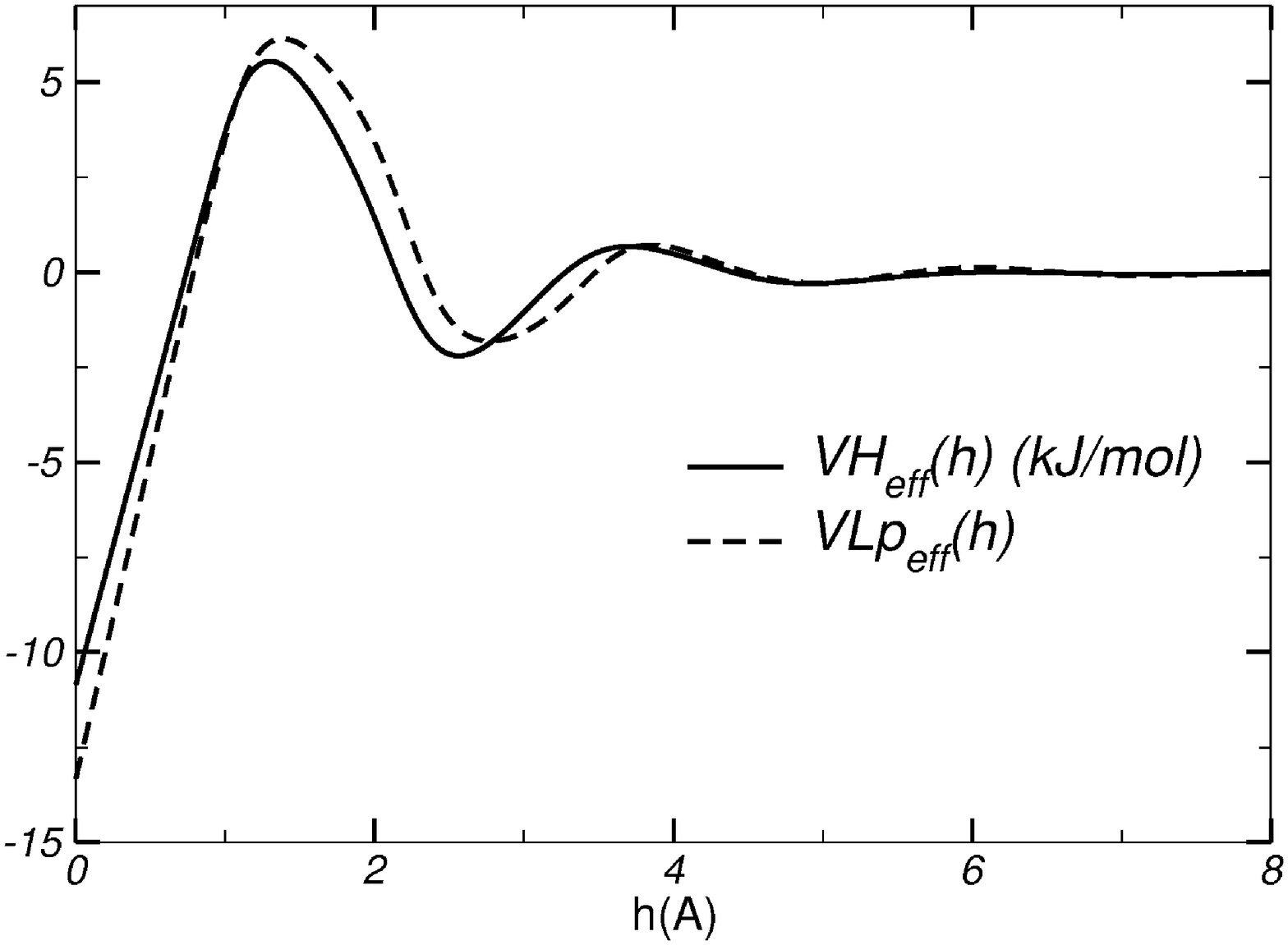}

\caption{The effective interaction potential on \emph{(a)} O (a LJ site) and
\emph{(b)} charged sites, located at a distance $h$ of the water
semi-volume.\label{cap:veff-0-qq}}
\end{figure}

The following figure, Fig. \ref{cap:veff-0-qq}(a) shows the interaction
energy between a LJ site at a distance $h$ of the liquid surface,
calculated with the pair correlation function $g_{OO}(r)$ and, for
the sake of comparison, we include the same interaction potential
with a homogeneous semi-volume \cite{steele}. Note that both potentials
greatly differ at short distances $z$. This fact is important, for
example, in studies on the interaction of confined water within hydrophobic
surfaces \cite{hydrophobic-surf}, where a large difference will be
found when performing the calculations using our effective interaction
potential, instead of that given by an uniform distribution of sites.
Fig. \ref{cap:veff-0-qq}(b) is the result obtained for q-q interactions,
calculated with the pair correlation functions $g_{HH}(r)$, $g_{LpLp}(r)$
y $g_{HLp}(r)$. The obtained $U_{eff}(h)$ functions closely follows
those measured in our MD sample of pure water at STP. \\

\section{Our amphiphilic molecule model solved in water}

In a next step we studied the pair distribution functions of an amphiphilic
molecule solved in water. We propose two simple amphiphilic models,
one is charged and the other is neutral, with a dipolar head. \\
Our charged amphiphilic consists in a semiflexible single chain of
14 atoms, the bond lengths are held constant, but bending and torsional
potentials are included. The first two atoms of the chain mimic a
charged and polar head (atom 1 with $q_{1}$ =- 2 e, atom 2 with $q_{2}$
= 1 e), sites 3 to 14 form the hydrophobic tail with uncharged atoms:
sites 3 to 13 are united atom sites $CH_{2}$ and site 14 is the united
atom site $CH_{3}$. This charged model is the same used in the simulation
of NB films \cite{zg-bub08} and correspond to an oversimplification
of sodium dodecyl sulfate SDS ($CH_{3}(CH_{2})_{11}OSO_{3}^{-}\, Na^{+}$)
in solution, so we are including  a $Na^{+}$ ion per chain. The LJ
parameters are those of ref. \cite{zg}, except for the sites 1 and
2 that form the amphiphilic polar head: $\sigma_{1}=4.0$$\textrm{Å}$,
$\sigma_{2}=$4.0 $\textrm{Å}$, $\sigma_{Na}=$1.897 $\textrm{Å}$,
$\varepsilon_{1}$= 2.20 kJ/mol, $\varepsilon_{2}$= 1.80 kJ/mol and
$\varepsilon_{Na}$= 6.721 kJ/mol. The masses of the sites are the
corresponding atomic masses, except that $m_{1}=m_{2}=48$au., in
order to mimic the 'real' amphiphilic head. The LJ parameters of the
united atom sites are taken from calcutations on \emph{n-}alkanes
\cite{pot.toxvaerd}: $\sigma_{CH2}=$3.850 $\textrm{Å}$, $\sigma_{CH3}=$3.850
$\textrm{Å}$, $\varepsilon_{CH2}$= 0.664 kJ/mol, $\varepsilon_{CH3}$=
0.997 kJ/mol. The LJ parameters for the $Na^{+}$ion are taken from
simulations of SDS in aqueous solution \cite{mike-sds-miscelle} and
Newton black films \cite{zg}. \\

The intramolecular potential includes harmonic wells for the bending
angles $\beta$ and the usual triple well for the torsional angles
$\tau$, the constants are the those commonly used for the united
atom site $CH_{2}$\cite{tildesley}. These potentials are needed
to maintain the amphiphilic stiffness and avoid molecular collapse.
The bending potential is\[
V(\beta)=k_{CCC}(\beta-\beta_{0})^{2}\;,\]
with $\beta_{0}=109.5\, deg.$ and $k_{CCC}=520\, kJ/rad^{2}$. The
torsional potential is of the form

\[
V(\tau)=a_{0}+a_{1}cos(\tau)+a_{2}cos^{2}(\tau)+a_{3}cos^{3}(\tau)+a_{4}cos^{4}(\tau)+a_{5}cos^{5}(\tau)\;,\]
the constants are $a_{0}=9.2789,$ $a_{1}=12.1557,$ $a_{2}=-13.1202,$
$a_{3}=-3.0597,$ $a_{4}=26.2403$ and $a_{5}=-31.4950\, kJ/mol$;
this potential has a main minimum at $\tau=0\, deg.$ and two secondary
minima at $\tau=\pm120\, deg.$\\
The neutral amphiphilic molecule is entirely similar to the charged
one, except that, in this case is $q_{1}=-q_{2}$ =- 1 e, and no ions
are included in the simulation.\\
\\

\begin{figure}
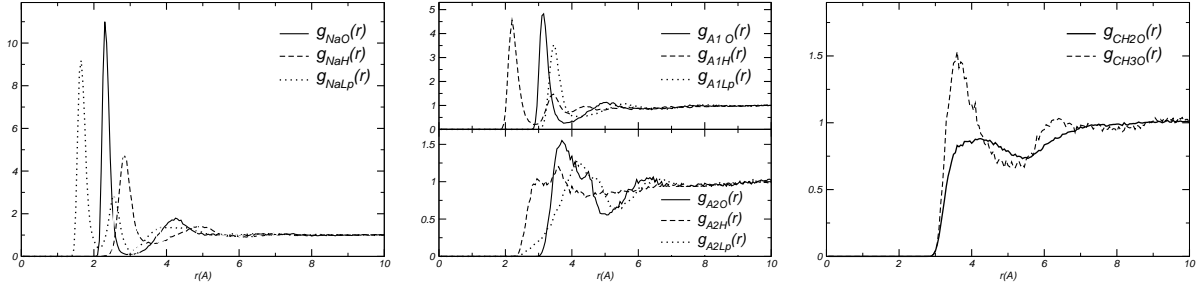

\includegraphics[clip,scale=0.23]{g2-w10iamp-na-mejor}$\quad$\includegraphics[clip,scale=0.23]{g2-w10iamp-a1a2-mejor}$\quad$\includegraphics[clip,scale=0.23]{g2-w10iamp-ch2ch3-mejor}

\caption{Pair correlation functions between the charged amphiphilic model
sites, unbonded ion and water sites.\label{cap:g2-ampiw} }
\end{figure}

\begin{figure}
\vspace{1cm}\includegraphics[clip,scale=0.22]{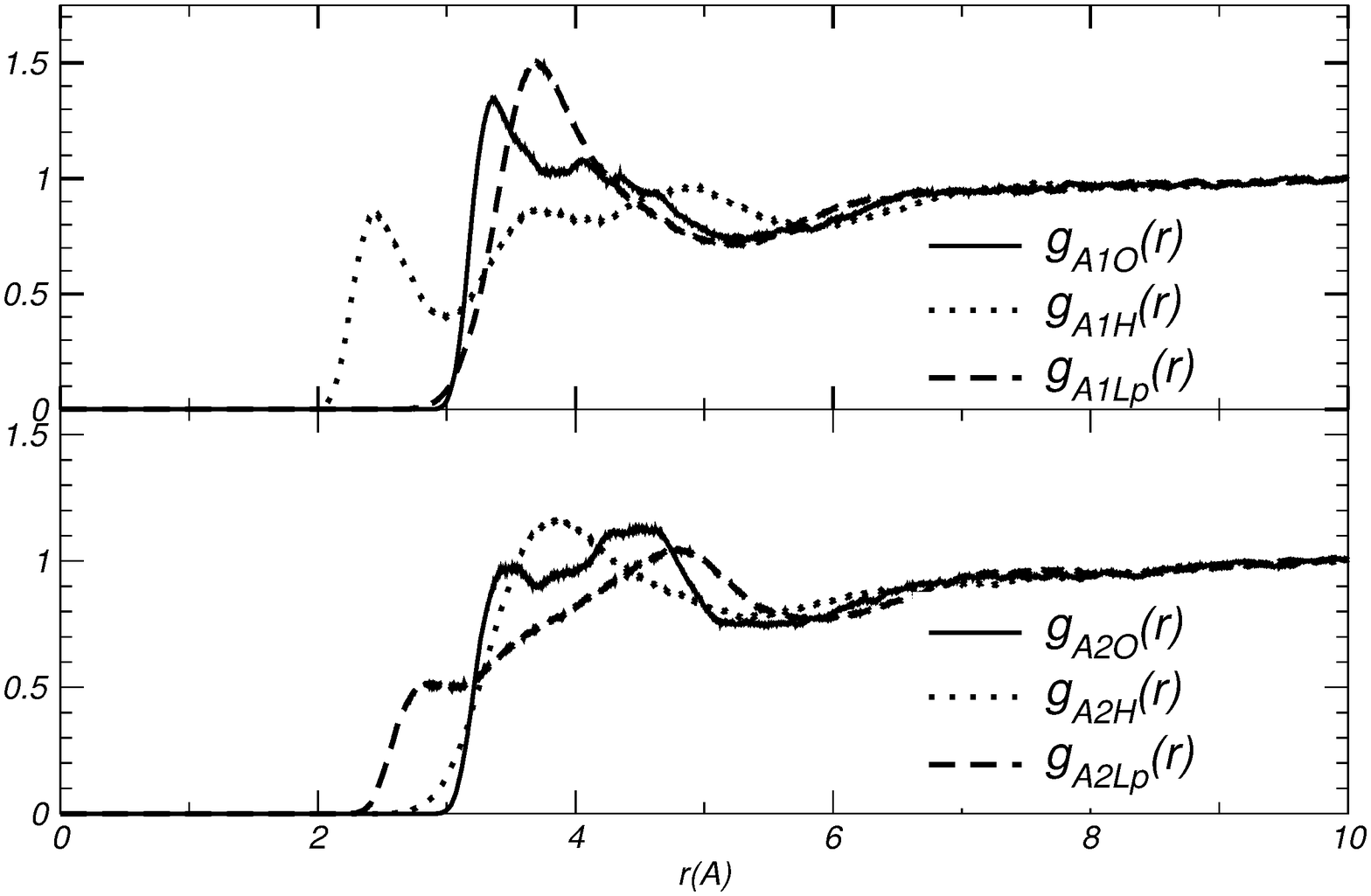}$\quad\quad$\includegraphics[clip,scale=0.22]{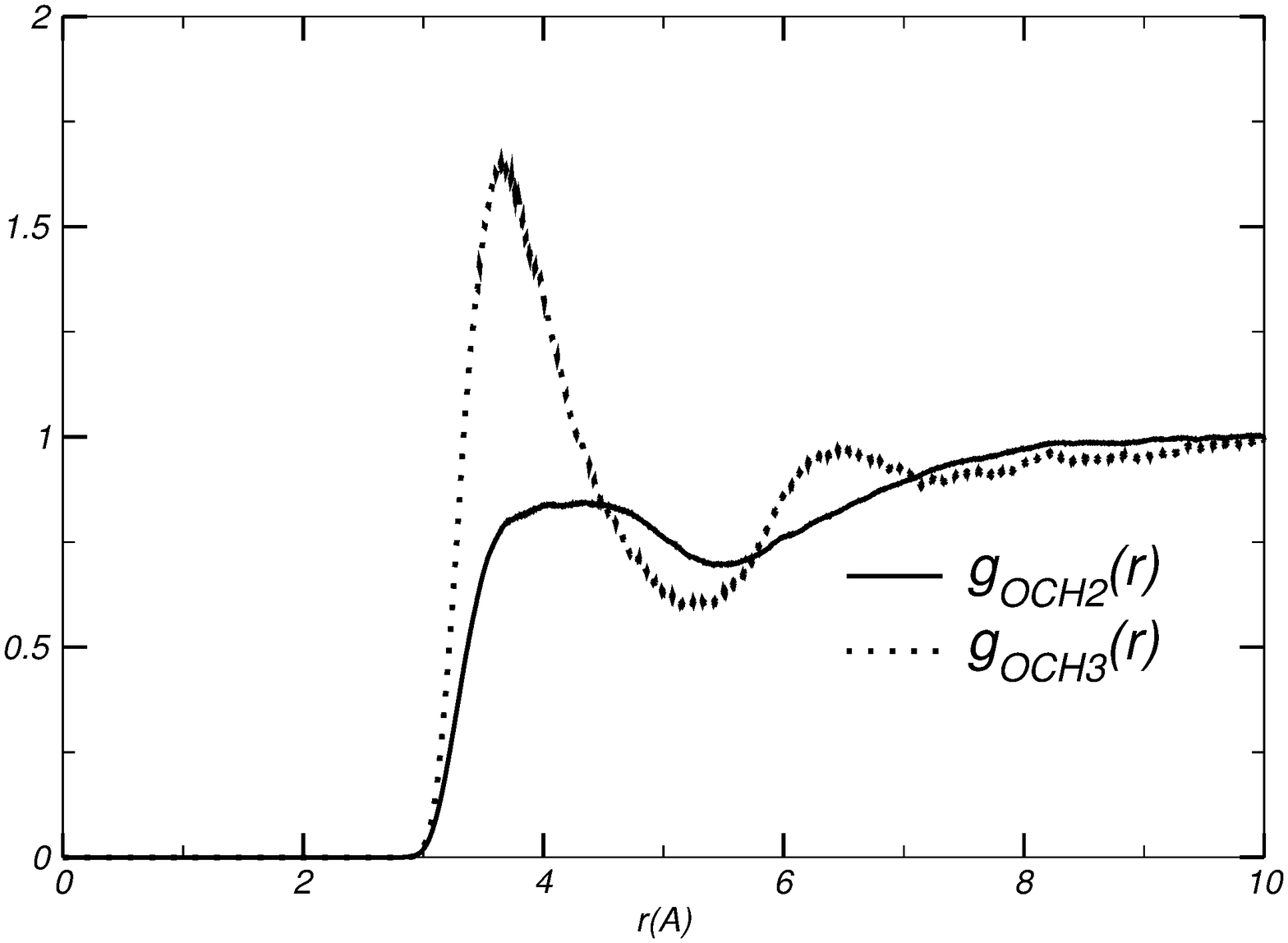}

\caption{Pair correlation functions between the neutral amphiphilic model
sites and water sites.\label{cap:g2-ampw}}
\end{figure}

\begin{figure}
\includegraphics[bb=41bp 48bp 691bp 535bp,clip,scale=0.3]{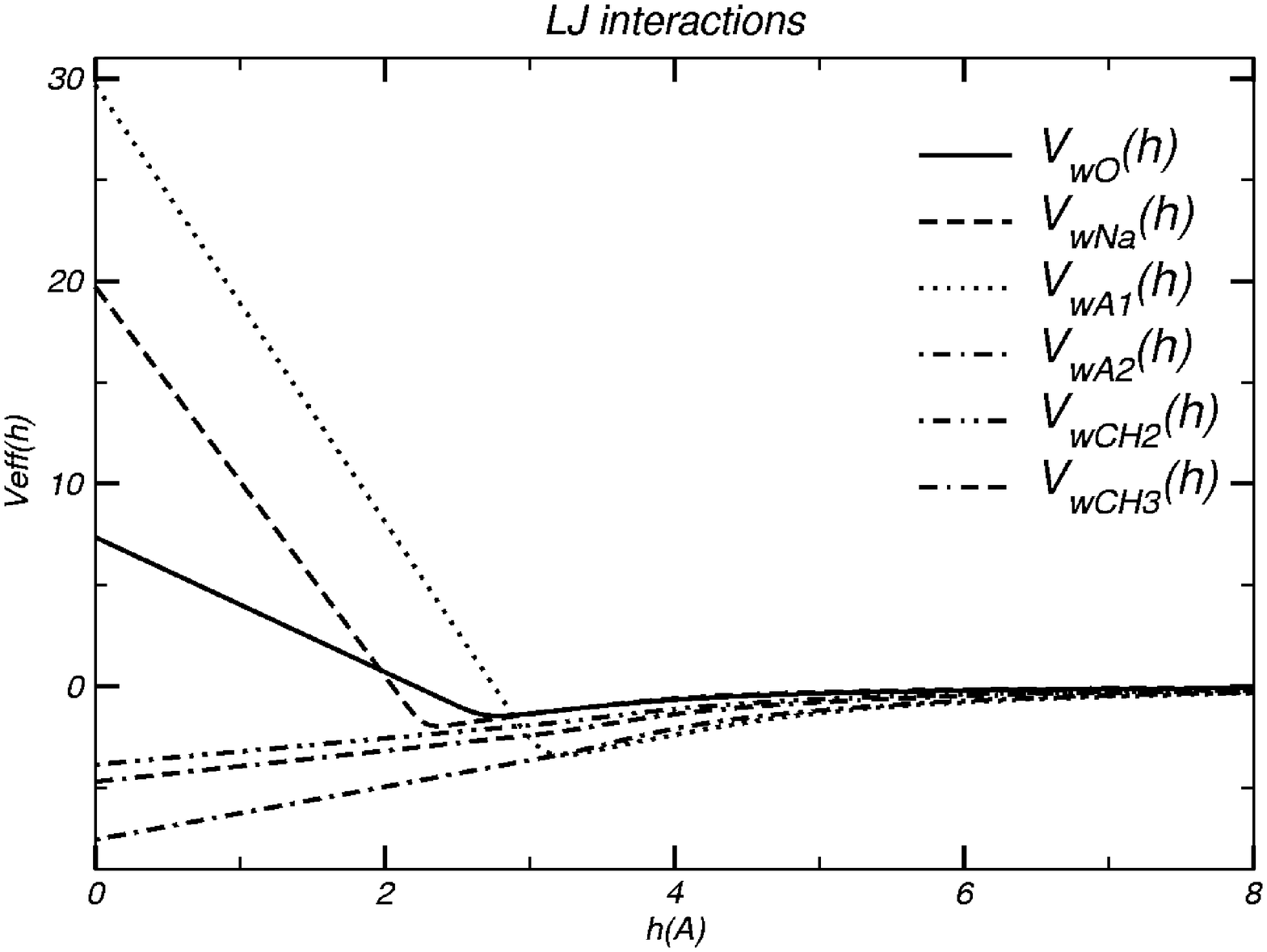}$\quad\quad\quad$\includegraphics[bb=44bp 48bp 691bp 508bp,clip,scale=0.3]{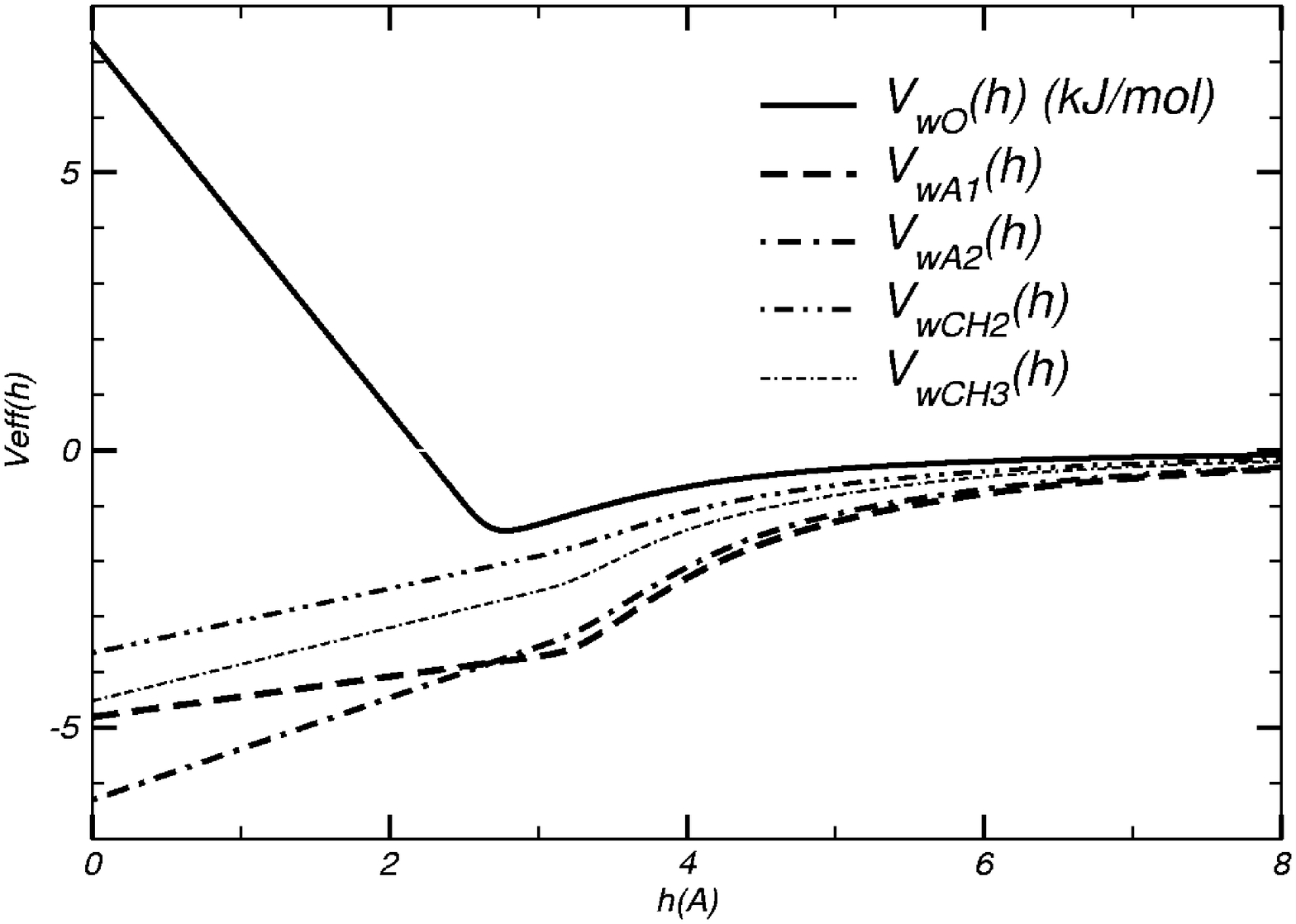}

\caption{Effective potentials for LJ sites of \emph{(a)} charged and \emph{(b)}
neutral amphiphilic sites interacting with the {}``liquid wall''.
Units as in Fig. \ref{cap:MD-eff-pf}\label{cap:feff-aa-amph}}
\end{figure}

\begin{figure}
\vspace{1cm}\includegraphics[scale=0.3]{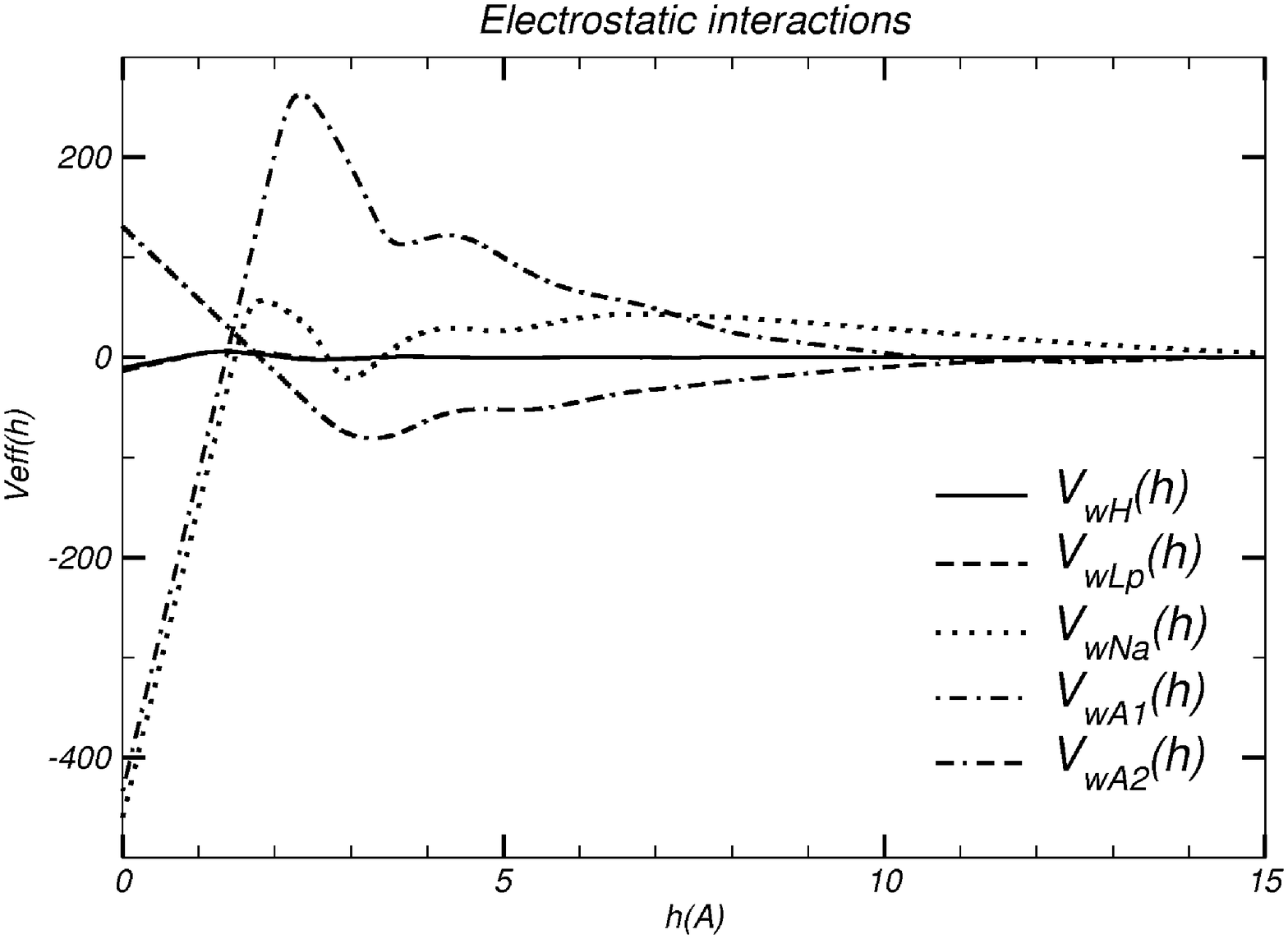}$\quad\quad\quad$\includegraphics[clip,scale=0.3]{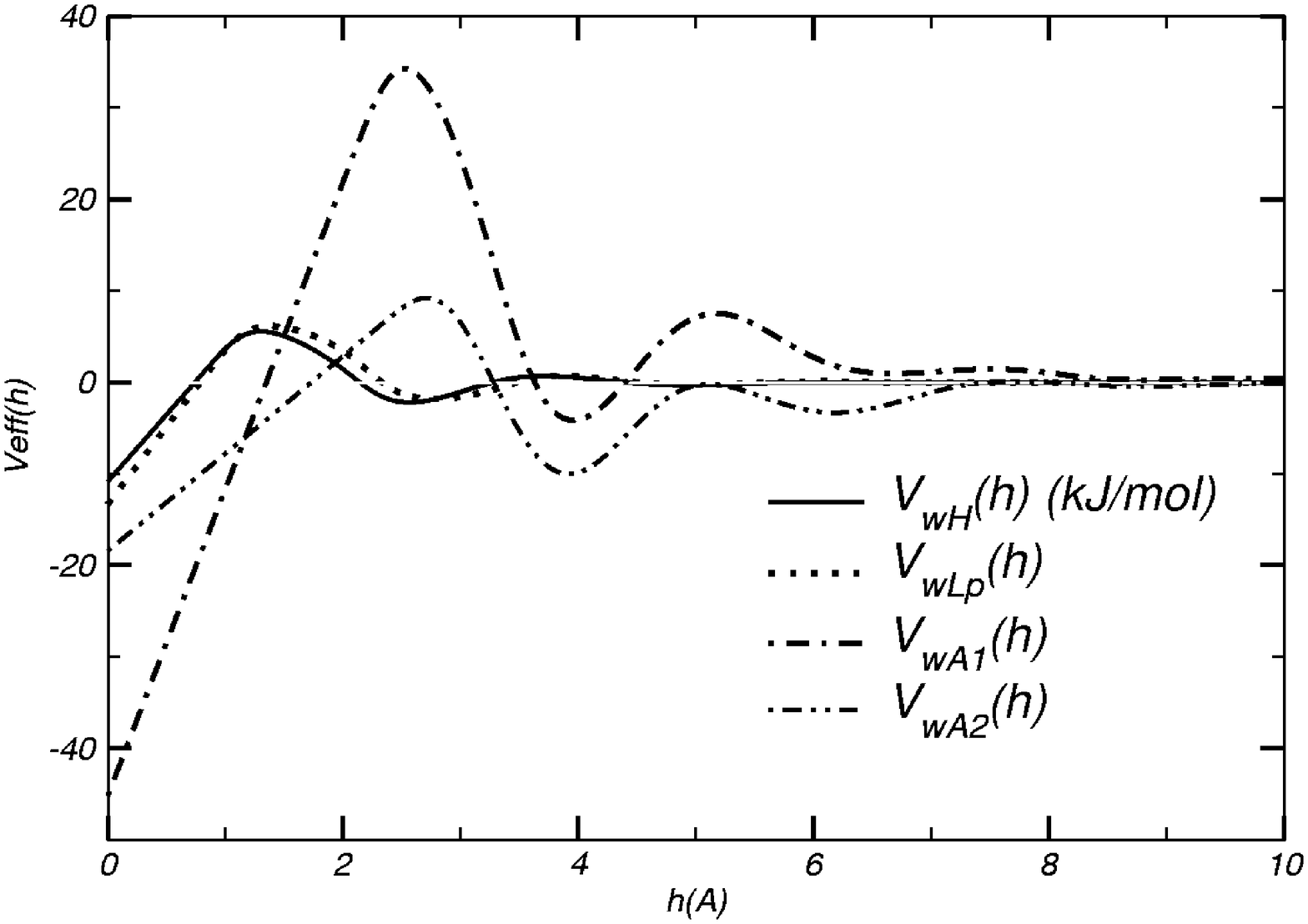}

\caption{Effective potentials calculated for charged sites of \emph{(a)} charged
and \emph{(b)} neutral amphiphilics interacting with the {}``liquid
wall''. Units as in Fig. \ref{cap:MD-eff-pf}\label{cap:feff-qq-amph}}
\end{figure}

The pair correlation functions between the amphiphilic and the TIP5P
sites were obtained from a MD simulation, performed with a time step
of 1 fs, 40 ps. of equilibration, and afterwards measured over a free
trajectory of 40 ps. The following figures include the pair correlation
functions (Fig.\ref{cap:g2-ampiw} for charged amphiphilics and Fig.
\ref{cap:g2-ampw} for neutral ones) and the corresponding effective
potentials for LJ sites, (Fig. \ref{cap:feff-aa-amph}) and charged
sites (Fig. \ref{cap:feff-qq-amph}), calculated as in the preceding
section for both amphiphilic models, interacting with a 'liquid semi-volume'.
\\
\\

\section{The biological bilayer model:}

\begin{figure}
[!ht]\includegraphics[width=0.18\paperwidth]{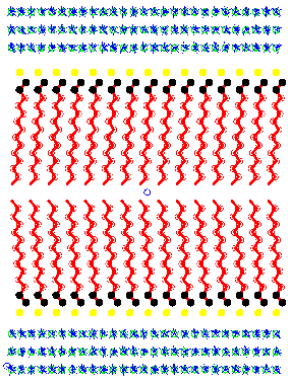}

\caption{Initial configurations of: a model biological membrane.\label{cap:mem226c14-iniconf}}
\end{figure}

Using the above mentioned molecular models we can build several simple
models of biological membranes and also a Newton black film \cite{zg-bub08}.
The initial configuration of our MD simulations is always a pre-assembled
structure, because the time scale of self-assembly, starting from
a homogeneus mixture of lipids and water, is about ns. \cite{mem-mike-coarse.grain},
the order of our longest simulations. For example, Fig. \ref{cap:mem226c14-iniconf}
shows the initial configuration of one of our simple biological membrane
simulations, consisting of 226 of our charged amphiphilics, 226 $Na^{+}$
ions and 2188 TIP5P molecules, the bilayer is perpendicular to the
$\widehat{z}$ MD box axis, with a box size of $a=b=45.$$\textrm{Å}$,
$c=$1000 $\textrm{Å}$. Although real biological membranes are usually
modeled with amphiphilic molecules consisting in one head with two
hydrophobic chains \cite{mem-mike-all.atom1}, here we are analysing
an extremely simple membrane model in which the strong electrostatic
interactions and the density of hydrophobic chains in the bilayer
play the key rôle. That is our reason for using the same amphiphilic
model as that in our NB films simulations \cite{zg-bub08}.\\

The MD integration algorithms, time step and cut-off radius are essentially
identical to those used on our bulk samples, except for the 2D periodic
boundary conditions, that now are applied only in the xy plane of
the bilayers, and that we are including: (a) an effective external
potential for biological membranes (in the direction perpendicular
to the bilayer) due to the surrounding water not explicitly included
in the simulation, and (b) our proposed macroscopic electric field.
\\

For our constant temperature MD simulations of bilayers we use the
Berendsen algorithm \cite{berendsen1}, applying the equipartition
theorem to each type of molecule and a strong coupling constant $\tau=0.5\,\Delta t$
linking the average kinetic energy of each kind of molecules (amphiphilics,
ions and water) to the desired kinetic energy of $\frac{3}{2}k_{B}T_{0}$,
with $T_{0}=300K$. The Berendsen algorithm turned out to attain equipartition
and equilibrium temperatures faster than the Nose-Hoover chains method
\cite{nose-chains} (used in our bulk samples), when applied to our
mix of flexible and rigid molecules.\\

We also have to take into account the following facts:

a) Although the $c$ parameter of the MD box is constant, at equilibrium
the width of the slab (that includes the layers of amphiphilic and
water molecules) fluctuates in time, mantaining a given average perpendicular
pressure, which we set about 1 atm in all included simulations. If
desired, by allowing variations of the \emph{ab} MD box size, the
lateral tension of the bilayer can also be adjusted to fluctuate around
a given value.

b) To simulate a \emph{single} biological membrane, all effective
interactions between our molecules included in the sample and those
water molecules outside the slab, we include \emph{effective potentials}
$U_{eff}(h)$, their contribution to the energy and forces on every
site depending on the type of site and the distance of the site to
both surfaces that delimit the slab. That means that at each time
step we need to determine the two distances $h^{-}$ and $h^{+}$,
of all molecular sites  to the instant location of the two confining
liquid surfaces. It is interesting to note that these effective potentials
and forces tend to maintain a tight packing of the bilayer along $\widehat{z}$.
This last statement was verified by running a sample similar to that
of Fig. \ref{cap:mem226c14-iniconf} \emph{(a)}, except that all ions
and amphiphilic molecules are removed. In a few ps. the two slabs
of water join in a single layer, with the final (and fluctuating)
thickness corresponding to the experimental STP density of water.

c) The site-site LJ interactions between all molecules in the sample
have a finite cutt-off radius of 15 A. In 3D simulatons, the contributions
of sites outside this sphere are taken into account assuming an uniform
distribution of sites and performing a simple integration. In our
cuasi 2D system, the volume to integrate is that outside the cut-off
sphere and within the volume of the slab. Appendix A in Ref. \cite{zg-bub08}
includes this integral.

d) The electrostatic interactions are calculated \emph{via} the standard
3D Ewald sums with a large size box along the perpendicular to the
bilayer slab and 2D periodic boundary conditions in the plane of the
slab. The Ewald's sum term corresponding to our proposed macroscopic
electric field is discussed in the following section.

\section{The macroscopic electric field in a quasi - 2 D sample:}

Electrostatic forces have a far from negligible contribution to the
self-assembly and final patterns found in soft matter systems. Several
reviews for quasi-2D and 3D geometries \cite{ewald-2d-deleeuw,jorge-2d,ewald-mz}
are available, where the macroscopic electric field is given, in a
first approximation, by the first multipolar (dipole) moment of the
MD box. In particular, for monolayers, the macroscopic electric field
is given in a first approximation by the contribution of the surface
charges of an uniform dielectric slab. If the slab, of volume $V$,
is oriented perpendicular to the z direction, the contribution to
the total energy of the system is:

$U^{macrosc.}=\frac{2\pi}{V}M_{z}²$, \\
where $M_{z}$ is the total dipole moment of the slab, and the contribution
of this term to the total force on every charge $q_{i}$ of the sample
is:

$F_{i}^{macrosc.}(z)=-\frac{4\pi}{V}M_{z}$.\\
This approximation for the macroscopic field, plus 3D Ewald sums with
a large MD cell along $z$, has been tested in simulations of monolayers
\cite{ewald-mz2,ewald-mz}. In the MD simulation of bilayers, instead,
and due to its geometry, the total dipole moment $M_{z}$ is zero
in a time average and therefore a more accurate estimation of their
macroscopic electric field is desirable. \\

In a recent paper \cite{zg-bub08} we discussed several approaches
and proposed a novel coarse fit of the charge distribution of the
different membrane components (water and amphiphilics plus ions),
using a superposition of gaussian distributions along $z$. In this
way, the contribution of these charge distributions to the macroscopic
electric field can be exactly calculated. The method is extremely
simple to implement in numerical simulations, and the spatial and
temporal charge inhomogeneities are roughly taken into account.

At each time step of the MD simulation we decompose our bilayer's
charge distribution in four neutral slabs: two for the upper and lower
water layers and other two for the amphiphilic heads plus ions. For
each one of the four neutral slabs, instead of consider two planar
surfaces with an uniform density of opposite charges, we propose two
gaussian distribution along \emph{z}, the perpendicular to the slabs,
with the same opposite total charges and located the same relative
distance (maintaining the slab width). The coarsed distribution of
charges in the bilayer is then a linear superposition of gaussians:\\

$\rho(z)=\sum_{i}\,\frac{q_{i}}{\sqrt{2\pi}\sigma_{i}}\exp(-\frac{(z-z_{i})²}{2\sigma²_{i}}))\quad,$\\

The macroscopic electric potential $V(z)$ and the force field $E_{z}(z)$
due to this type of charge distribution can be exactly solved. In
Appendix C of Ref. \cite{zg-bub08} we include the analytically solved
integrals (one of them is a new integral not included in Mathematica
\cite{math}). The final result is:\\

$V(z)=-2\sqrt{2\pi}\sum_{i}\,\sigma_{i}\, q_{i}\,(\exp(-\frac{(z-z_{i})²}{2\sigma_{i}²})-(\frac{z-z_{i}}{\sqrt{2}\sigma_{i}})\, Erf[\frac{(z-z_{i})}{\sqrt{2}\sigma_{i}}])\,.$

$E_{z}(z)=-\frac{\partial V(z)}{\partial z}=2\pi\sum_{i}\, q_{i}\, Erf[\frac{(z-z_{i})}{\sqrt{2}\sigma_{i}}].$\\

These expresions are valid for any number of slabs, that as a function
of time can change not only their position and width but also they
can superpose. To include the macroscopic electric field term we need
to determine the values of the $\sigma_{i},$ $z_{i}$ and $q_{i}$
parameters at each MD time step. For each one of the two slabs that
simulate the charge distribution of chains' heads plus ions, we fit
the $z_{i}$ parameters of two gaussians, so as to reproduce the dipolar
moment of the slab. Their $\sigma_{i}$ values are obtained from the
corresponding charge distributions, with $q_{i}=1$ for ions and $q_{i}=-1$
for chains' heads. Typical values of these variables, as well as the
contribution of the external potential and the macroscopic electric
field to the total forces on all molecules, are reported for all simulated
bilayers in section VI. \\

Here we have applied this exact calculation method of the macroscopic
electric field to a symmetrical (along \emph{z}) slab geometry, but
it is also valid in an asymmetrical case, which may imply a finite
difference of potential across the bilayer. As we pointed out in Ref.
\cite{zg-bub08}, the extension of this method ( a coarse grained
representation of the macroscopic electric field \emph{via} a superposition
of gaussians) to other geometries is also strightforward, a spherical
geometry, for example, would be useful for the study of miscelles.
Its great advantage is that these representations can be a\emph{nalytically}
solved. \\

\section{Four calculated model bilayers, results:}

To test the versatility of our approach, four different bilayers models
were studied: with and without ions solved in water and with and without
the water layer. We only include here a few simulations of each simplified
bilayer model, which were mainly performed to analyze the contribution
of the effective external potential and the macroscopic electric fields
to the equilibrium structure and molecular dynamics of each bilayer.
Elsewhere we will present a detailed analysis of the phase diagram
of these and other bilayer models and their dependence on the amount
of water and ions, the amphiphilics length, external pressure, lateral
tension, etc.

\subsection{A simple biological membrane with its surrounding water and solved
ions: }

\begin{figure}
[!ht]\includegraphics[scale=0.4]{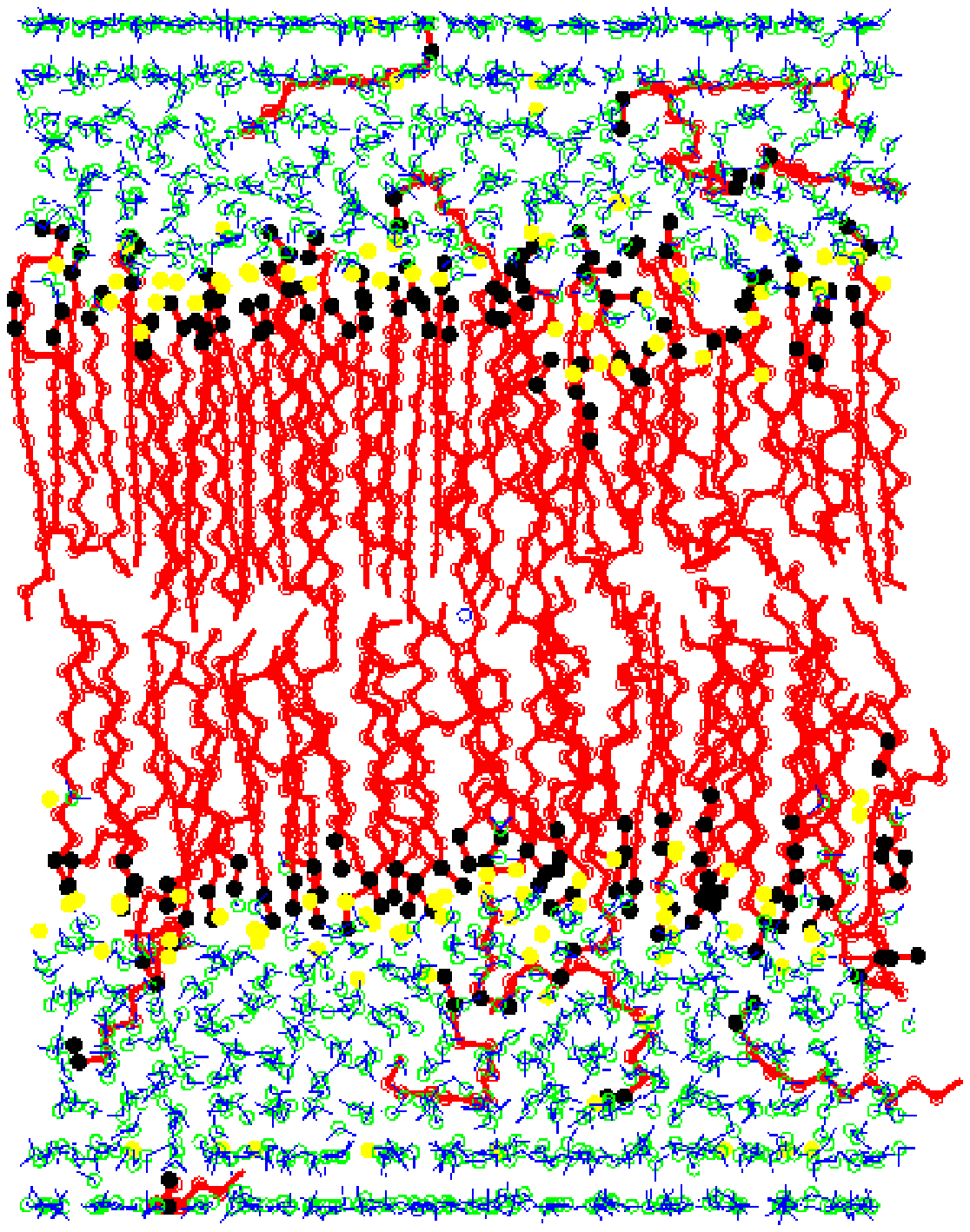}$\qquad$\includegraphics[scale=0.4]{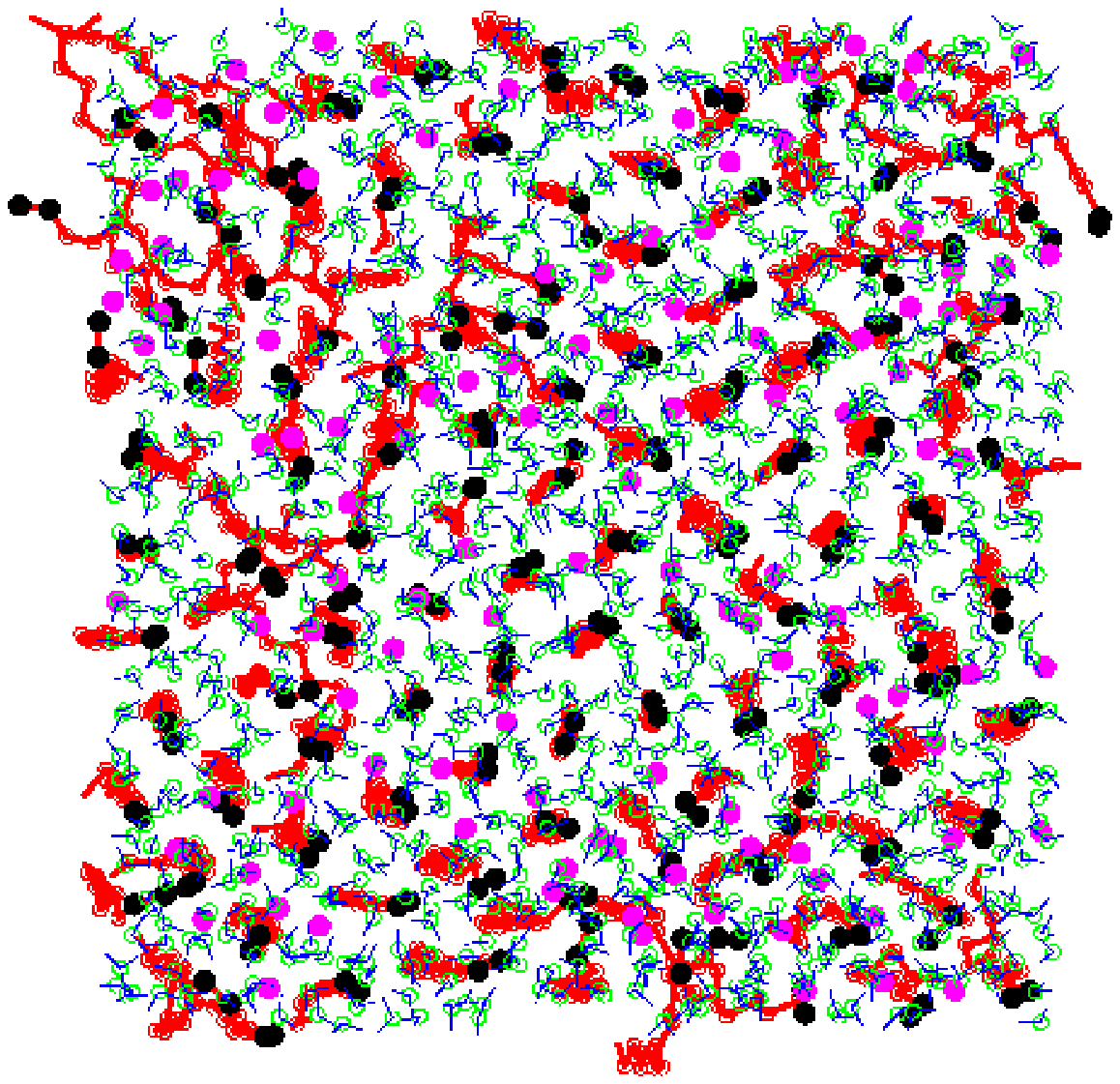}

\caption{Final configuration of the sample of Fig. \ref{cap:mem226c14-iniconf}a,
with \emph{a}=\emph{b}=42.45$\textrm{Å}$: (a) \emph{ac} cross section,
(b) \emph{ab} cross section. \label{cap:Final-conf-226c14-chiw}}
\end{figure}

\begin{figure}
[!ht]\includegraphics[scale=0.41]{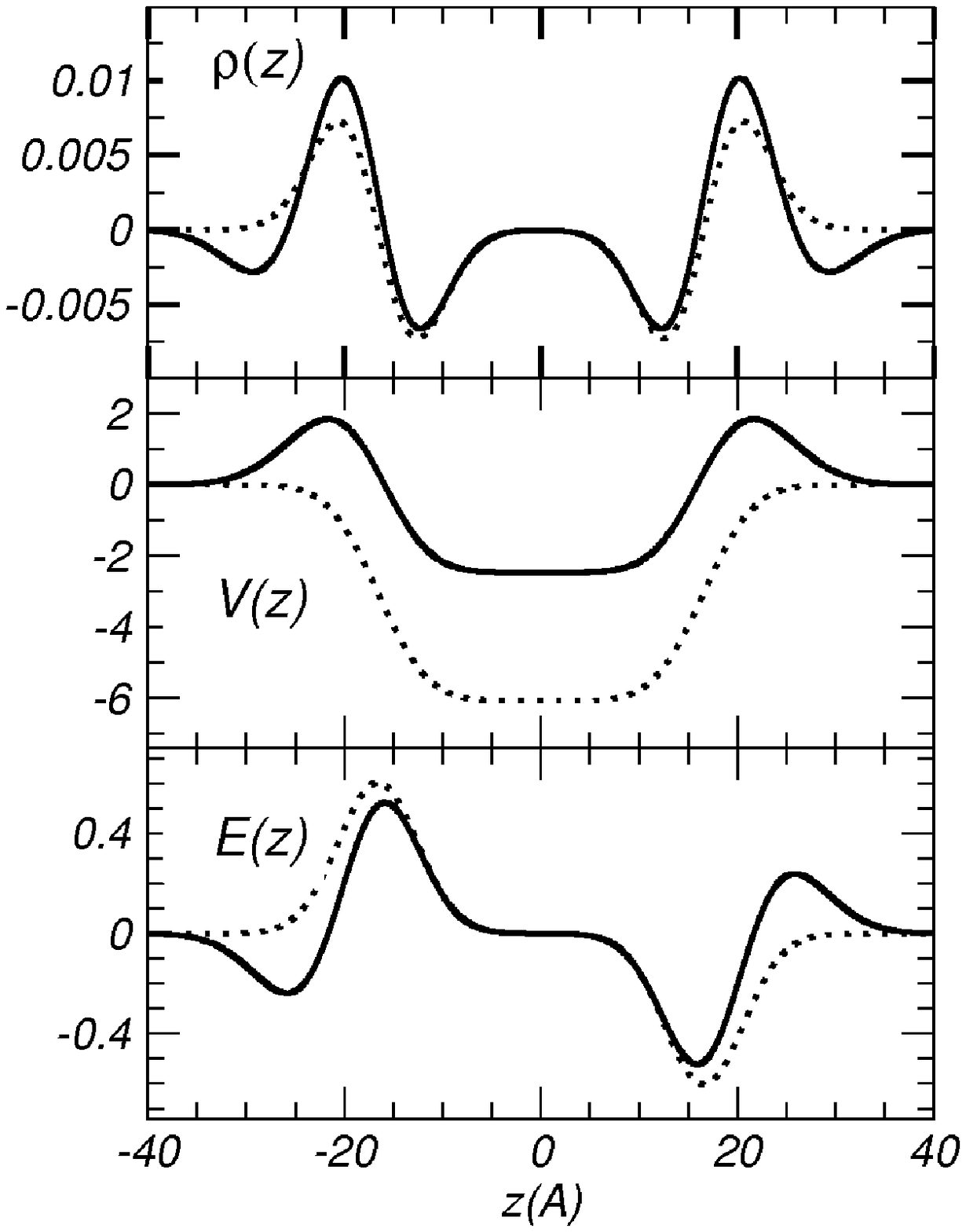}$\qquad\qquad\qquad\qquad$\includegraphics[scale=0.41]{mems-226c14-chiw-a42\lyxdot 45-a8d-fzmido-prom}

\caption{Profile of: (a) the macroscopic coarse fit of the charge density
$\rho$(z), the macroscopic electrostatic potential \emph{V(z)} and
electric field \emph{E(z)} (full line for total values, dotted line
for heads plus ions contribution), (b) our measured profile of forces
acting on all molecules due to the macroscopic electric field, the
external force and total forces, as a function of the centers of mass
locations along $z$, in the model membrane with charged chains. \label{cap:Prof.Emacro-fz-226c14-cchiw-HD}}
\end{figure}

This is the sample whose initial configuration was included in section
IV. Fig. \ref{cap:mem226c14-iniconf}\emph{(a)} shows the initial
and Fig. \ref{cap:Final-conf-226c14-chiw} the final equilibrated
configuration of this simple model membrane, consisting of 226 negatively
charged amphiphilic molecules, 226 positive charged ions and 2188
water molecules (9.7 water molecules per amphiphilic), the bilayer
is perpendicular to the $\widehat{z}$ MD box axis, with a box size
of $a=b=42.45$$\textrm{Å}$, $c=$1000 $\textrm{Å}$. In a constant
volume MD simulation the final equilibrium width of this slab (including
all molecules) is about 60$\textrm{Å}$. The periodic boundary conditions
are applied along the $\widehat{x}$ and $\widehat{y}$ directions
and a large unit cell $c$ parameter (that is, a large empty volume)
is taken, in order to approximate the required 2D Ewald sums by the
usual 3D sums. In addition, the contribution of the effective external
potential due to the water outside the slab and the macroscopic electric
field are taken into account (Fig. \ref{cap:Prof.Emacro-fz-226c14-cchiw-HD}).\\
\vspace{1cm}

Fig. \ref{cap:Prof.Emacro-fz-226c14-cchiw-HD} \emph{(a)} shows the
functions $\rho(z)$, $V(z)$ and $E_{z}(z)$ calculated in one of
the time steps of the free MD trajectory of the equilibrated sample.
As time evolves, these functions show small fluctuations around the
values included in the figure. $\rho(z)$ is obtained from a coarse
grained fit of the charge distribution of our MD sample using four
charged slabs, two of them represent the charged heads plus ions and
the other two the water molecules (As explained in section V). The
parameters of the four slabs used to calculate $\rho(z)$ in Fig.
\ref{cap:Prof.Emacro-fz-226c14-cchiw-HD} are:

$\begin{array}{lrrr}
 & q_{i}(e) & z_{i}(\textrm{Å}) & \sigma_{i}(\textrm{Å})\\
slab\,1\,(water) & -0.241 & 24.515 & 4.705\\
 & 0.241 & 23.325 & 4.705\\
slab\,2\,(head\, and\, ions) & -1.0 & 16.349 & 4.011\\
 & 1.0 & 16.832 & 4.011\\
slab\,3\,(head\, and\, ions) & 1.0 & -16.832 & 4.011\\
 & -1.0 & -16.349 & 4.011\\
slab\,4\,(water) & 0.241 & -23.325 & 4.705\\
 & -0.241 & -24.515 & 4.705\end{array}$. \\

Once obtained $\rho(z)$, the macroscopic potential $V(z)$ and corresponding
electric field $E_{z}(z)$ are calculated as explained in section
V. Units in Fig. \ref{cap:Prof.Emacro-fz-226c14-cchiw-HD} and following
ones: density of charges $\left[\rho\right]=\frac{e}{\textrm{Å}^{3}}$;
electrostatic potential $\left[V\right]=\frac{e}{\textrm{Å}}$ (for
comparison with experimental data $\left[\frac{e}{\textrm{Å}}\right]=0.04803\,\frac{cm^{1/2}gr^{1/2}}{sec}=14.399\,\left[Volt\right]$
); electric field $\left[E\right]=\frac{e}{\textrm{\textrm{Å}}^{2}}$.\\

\begin{figure}
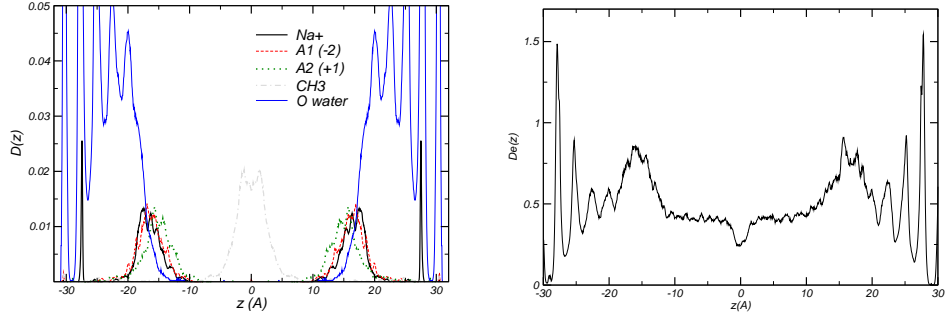

\includegraphics[scale=0.25]{mems-226c14-chiw-a42\lyxdot 45-a8d-otra\lyxdot densa\lyxdot angstrom-color}$\qquad$\includegraphics[scale=0.25]{mems-226c14-chiw-a42\lyxdot 45-a8d-zdensel\lyxdot angstrom3}

\caption{\emph{(a)} Atomic (number of atoms per $\textrm{Å}^{3}$) and \emph{(b)}
electronic density profiles for the sample with \emph{a=b=}42.45$\textrm{Å}$\label{cap:At.dens.profiles-226c14-cchiw}.}
\end{figure}

Most of available all-atom simulations are on bilayers of neutral
amphiphilics. In our case, as the ion $Na^{+}$ is not bonded to the
amphiphilic head, we cannot directly compare our results. Our model
bilayer with charged amphiphilics is more similar to that calculated
in Ref. \cite{mem-faraudo-07}, where a bilayer of 100 $DMPA^{2-}$
(charge=-2e) amphiphilics, 9132 water molecules and $Cl_{2}Ba$ in
solution (100 $Cl^{-}$ and 150 $Ba^{2+}$ ions) is simulated. In
this sample they observe a charge inversion of about 1.07 Ba(2+) ions
per DMPA molecule, and therefore the contribution of lipids plus ions
to the electrostatic potential is positive in the core of the bilayer,
the contribution of the water layer is opposite but not large enough
so as to change the sign of the total potential. Our single chain
amphiphilic has a charge of $-1e$, but from the atomic density profiles
of Fig. \ref{cap:At.dens.profiles-226c14-cchiw} we calculate that
most of the ions $Na^{+}$ remain around the head group, but about
11\% of them are solved in the water layer. That means that our bilayer
do not show charge inversion and the contribution of lipids plus ions
to the electrostatic potential is negative, and we also observe that
the contribution of the water layer is opposite but not large enough
so as to change the sign of the total potential. We obtain for charged
head and ions a negative potential contribution of about -6 $e/\textrm{Å}$,
the water layer polarization is not strong enough to counterbalance
this trend, and the final value is a negative potential, at the membrane
core, of about -2 $e/\textrm{Å}$.\\

This behavior is at variance with the usual one observed for neutral
lipid bilayers, where a dielectric overscreening of water is observed
and therefore the total calculated electrostatic potential $V(z)$
is opposite to the contribution given by the amphiphilic layers. For
example, in a typical all-atom MD simulation of a neutral SDPC (1-stearoyl-2-docosahexaenoyl-sn-glycero-3-phosphocholine)
phospholipid bilayer \cite{mem-electr-mike-1}, it is found that,
on average, the negative P
 atom in the head group is located closer to the membrane interior
than the positive N atom, the same orientation for charges is obtained
with our coarse grained fit of the head groups plus ions. The SDPC
lipid dipoles \cite{mem-electr-mike-1} are oriented so as to contribute
with a negative electrostatic potential in the membrane core of about
-1 V ($\sim-14.4e/\textrm{Å}$), the water polarization creates an
opposite potential and the total result is a positive potential of
about +0.5 V \cite{mem-electr-mike-1}. A similar positive potential
at the membrane core of $\sim+0.35$V is calculated in an all-atom
MD simulation of a POPC (palmitoyl-oleoyl-phosphatidylcholine) lipid
membrane \cite{mem-POPC} and of +0.575V in other MD simulation of
a DPPC (dipalmitoy-phosphatidyl-choline) bilayer \cite{mem-DPPC-berendsen}.

Fig. \ref{cap:Prof.Emacro-fz-226c14-cchiw-HD} \emph{(b)} includes
the contribution of the effective external potentials $U_{eff}(z)$
(that models the surrounding water) and the macroscopic electric field
$E_{macro}(z)$ to our measured total forces on water, ions and chain
molecules, as a function of the center of mass locations along \emph{z},
averaged over a free MD trajectory of 50 ps, Units: $\left[F\right]=kJ/mol/\textrm{Å}$.
The negatively charged amphiphilics tend to drift away of the bilayer
due to the macroscopic electric field but this tendency is balanced
by an opposite force on the movile ions located in the neighborhood
of the charged heads. The external force due to the surrounding water
is mainly felt by the molecules near the up and lower borders of the
sample, although near the head groups follows the same tendency of
the forces due to the macroscopic field. Lastly, when adding all molecule
- molecule interactions, the total forces on the amphiphilics tend
to maintain a bilayer structure.\\

Figs. \ref{cap:At.dens.profiles-226c14-cchiw}\emph{(a)} and \emph{(b)}
show, respectively, our atomic and electronic density profiles. The
head to head distance, perpendicular to the bilayer, is about 36$\textrm{Å}$
with an area per hydrophobic chain of $16\textrm{Å}^{2}$. Our profile
for the water density is similar to those calculated for all-atom
samples only around the amphiphilic polar heads, but near both surfaces
of the slab, our profile resembles that calculated for water interacting
with a solid surface, although the fluctuations of density are smaller
due to a less repulsive short range external potential (Fig. \ref{cap:veff-0-qq}
\emph{(a)}). The last difference is due to our restriction on the
water diffusion along \emph{z} and to have allowed only very small
fluctuations of the internal pressure in the \emph{z} direction, this
in turn implies that the molecules near the up and lower border of
the sample, at less than 2-3 times the LJ size of water, should not
be taken into account when measuring the different bilayer properties.
\\

To obtain a fast overview of how these properties depend on the amphiphilics
density, we performed a small series of five constant volume - constant
temperature simulations, with increasing values of \emph{ab} cross
section, up to $a=b=45.0$$\textrm{Å}$, at which value large amounts
of water molecules interpenetrate the membrane and disorder the bilayer
structure, as measured by the diffusion coefficients and the atomic
density profile. In all simulations the pressure on the membrane,
along \emph{z}, is maintained fluctuating about 1atm., in the sample
of section (42.45 $\textrm{Å}$)\ensuremath{²} the lateral pressure
is about 3 atm, falling to $\sim$2 atm in the sample of section (45.0
$\textrm{Å}$)\ensuremath{²} , and the total width of the sample drops
from 60$\textrm{Å}$ to about 50$\textrm{Å}$. \\

The diffusion coeficients for the amphiphilic chains, ions and water,
are highly anisotropic in all samples. For the sake of comparison,
the experimental STP ($25^{\circ}$C and 1 atm.) diffusion coeficient
of bulk water \cite{w-dif} is $2.30\,10^{-5}cm^{2}/\sec$, and the
lateral diffusion of lipids in a stack of DPPC bilayers \cite{DPPC-dif},
for example, is about $\,10^{-6}cm^{2}/\sec$ at 330 K and less than
$0.6\,10^{-6}cm^{2}/\sec$ at 300 K. The next Table gives our measured
diffusion coeficients for amphiphilic chains, ions and water, in the
\emph{xy} plane and in the \emph{z} direction (Units: $10^{-5}cm^{2}/sec$).
It has to be taken into account that for simulation samples of our
size it is necessary to perform MD runs of nanoseconds to meassure
diffusion coeficients of the order of $10^{-6}cm^{2}/\sec$. As our
free MD trajectories are of 100 ps, our values are an upper limit
to the diffusion coeficients of these samples\cite{bilayer-lipid-diff}.
In the sample of section (42.45 $\textrm{Å}$\ensuremath{²}) we measured
oscillations of the centers of mass around an equilibrium position,
without diffusion, and correspond to a simulation of a glassy phase,
while for the sample of section (45.0 $\textrm{Å}$\ensuremath{²})
the measured difusion coefficients are of the order of $10^{-5}cm^{2}/\sec$
and correspond to a liquid phase of solved amphiphilics in water.\\

$\begin{array}{cllllll}
 & chain^{-}(z) & chain^{-}(xy) & ion^{+}(z) & ion^{+}(xy) & W(z) & W(xy)\\
a=42.45\textrm{Å}\quad & \:- & - & - & - & - & -\\
a=43.20\textrm{Å}\quad & \:0.2 & 0.7 & 0.15 & 0.6 & 0.2 & 1.2\\
a=43.50\textrm{Å}\quad & \:0.5 & 1.8 & 0.45 & 1.5 & 0.45 & 2.2\\
a=45.00\textrm{Å}\quad & \:1.6 & 3.8 & 1.6 & 3.1 & 1.0 & 3.8\end{array}$\\
\\

Finally, the distortion of the amphiphilic molecules was determined
by the fraction of \emph{trans} to \emph{gauche (g+, g- )} torsional
angles and by the distribution of site-site intramolecular distances
(Fig. \ref{cap:cchwi-head-tail-d}), calculated on a free MD trajectory
of 100 ps. Both measurements show an increasing chain disorder for
samples of lower density in the \emph{xy} plane.\\

\begin{floatingfigure}[l]{0.5\columnwidth}%
\vspace{-3cm}\includegraphics[scale=0.3]{mems-226c14-chiw-a42\lyxdot 45ya45\lyxdot 0-fdc}

\caption{The head to bead 4 and head to tail site distances distributions
for the amphiphilics in two samples. \label{cap:cchwi-head-tail-d}}\end{floatingfigure}%

$\begin{array}{ccc}
 & \quad trans & \; g+=g-\\
a=42.45\textrm{Å} & \:0.88 & 0.06\\
a=43.2\textrm{0Å} & \:0.88 & 0.06\\
a=43.50\textrm{Å} & \:0.86 & 0.07\\
a=45.00\textrm{Å} & \:0.66 & 0.17\end{array}$ \\
\vspace{7cm}

\subsection{A simple biological membrane without surrounding water:}

\begin{figure}
[!ht]\includegraphics[scale=0.4]{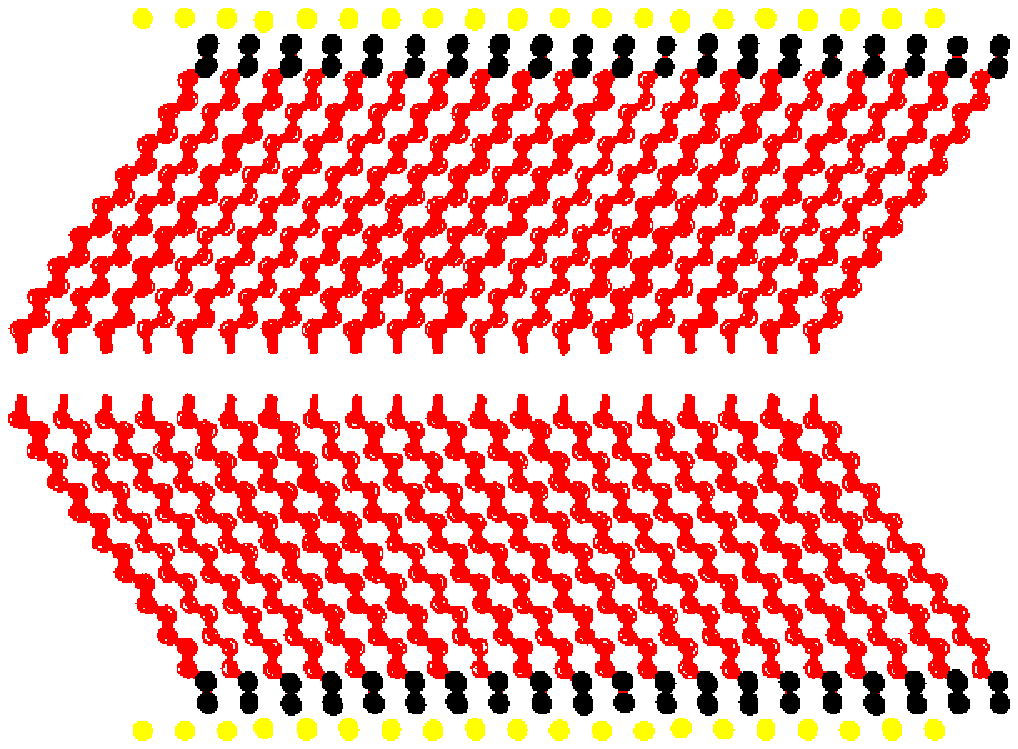}$\quad$\includegraphics[scale=0.4]{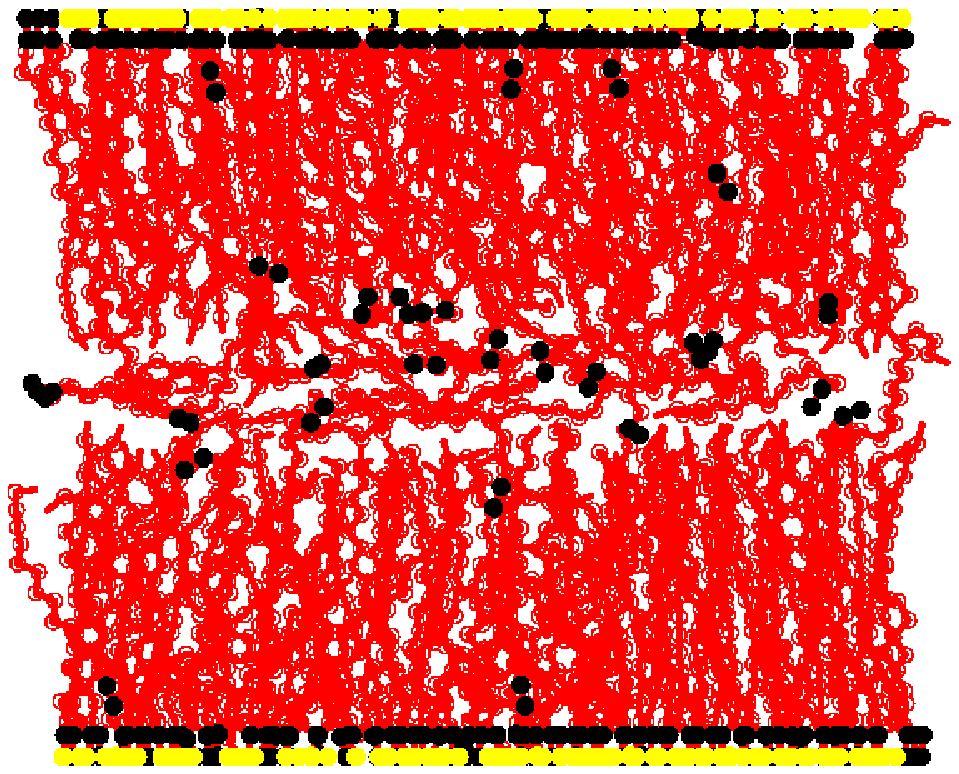}$\quad$\includegraphics[scale=0.44]{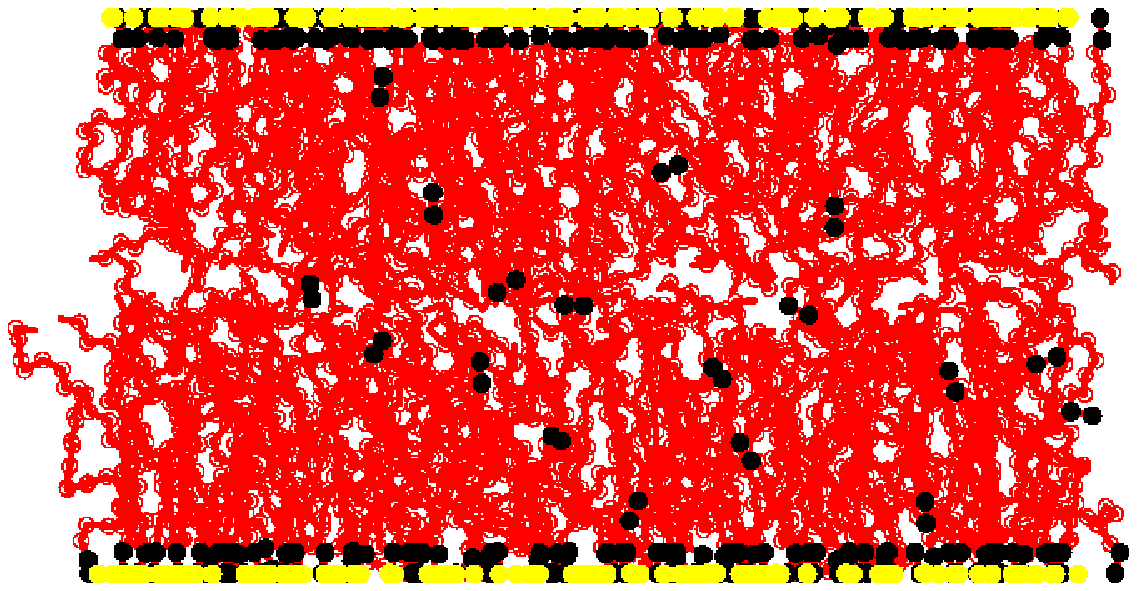}

\caption{Simple model membrane without water: (a) initial configuration, (b)
final configuration, (c) final configuration of a sample with lower
amphiphilics density.\label{cap:mem-400-conf}}
\end{figure}

In general, the models of amphiphilic membranes without the explicit
inclusion of the solvent (water) are investigated because of their
simplicity and faster calculation of fenomena at the mesoscopic scale,
as presented for example in Refs. \cite{farago,mem-potextra1,brannigan2}
where the self-assembly of bilayers is studied using simple neutral
chains of 3 to 5 beads, one of them representing the head. 

Here we present a model that is simple and useful at the nanoscopic
scale, that includes strong electrostatic interactions, flexible chains
of a more realistic number of atoms, and the LJ interaction parameters
for the hydrophobic tail are those used in all atom simulations of
gel or liquid phases of amphiphilics. All these facts make it a reliable
model to study the properties of guest molecules confined within a
biological membrane, making it possible to measure their diffusion
times, uncoiling dynamics of chain molecules, interaction with head
groups, etc. 

Here we include a simple model bilayer consisting on 400 negatively
charged chains plus 400 positive ions. We perform two MD simulations,
one with a box size $a=b=56.$$\textrm{Å}$ (HD sample with $16\,Å^{2}$per
chain), and a second one with $a=b=65$$\textrm{Å}$ (LD sample with
a density of $21\,Å^{2}$per chain), $c=$1000 $\textrm{Å}$ for both.
As an example of the versatility of our model, the chains in this
section consists in 20 beads, 2 of them forming the strongly charged
head. Except for their length, these chains are entirely similar to
those of the preceding section. As in the other cases included in
this chapter, the phase diagram of this simple membrane, and their
dependence on the chain lenght are currently under study. Fig. \ref{cap:mem-400-conf}(a)
shows the initial and \ref{cap:mem-400-conf}(b) the final configuration
of the HD sample, after a free MD trajectory of 100 ps, its equilibrium
thickness of 52$\textrm{Å}$. Fig. \ref{cap:mem-400-conf}(c) shows
the final configuration of the LD sample with an equilibrium thickness
of 40$\textrm{Å}$. \\

\begin{figure}
[ht!]\includegraphics[clip,scale=0.37]{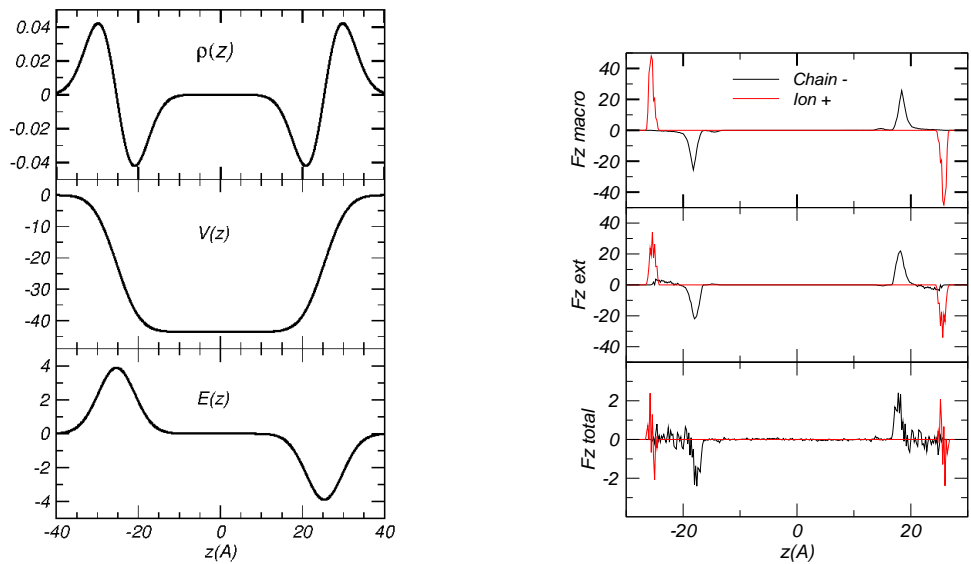}$\qquad\qquad\qquad$\includegraphics[clip,scale=0.42]{mems-chi-sinw-400c20-a56-alg25-a2f-fzmido-plot}

\caption{Profile of: (a) the macroscopic coarse fit of the charge density
$\rho$\emph{(z)}, potential \emph{V}(\emph{z}) and macroscopic electric
field \emph{E}(z) (full line for total values, dotted line for heads
plus ions contribution), (b) our measured profile of macroscopic electric
field, external force and total forces acting on all molecules, as
a function of the location of their centers of mass along $z$, in
the model membrane without water. \label{cap:Prof.fz-400chi}\protect \\
}
\end{figure}

The MD runs are performed including the macroscopic electric field
term and the external potential which simulates the surrounding water.
Fig. \ref{cap:Prof.fz-400chi} (a) shows the functions $\rho(z)$,
$V(z)$ and $E_{z}(z)$ calculated in one of the time steps in the
free trajectory of the equilibrated sample. In this case, $\rho(z)$
is obtained from a coarse grained fit of the charge distribution of
our MD sample using only two charged slabs that represent the charged
heads plus ions (As explained in section V). The parameters used to
calculate the functions in Fig. \ref{cap:Prof.fz-400chi} (a) are:
\\

$\begin{array}{lrrr}
 & q_{i}(e) & z_{i}(\textrm{Å}) & \sigma_{i}(\textrm{Å})\\
slab\,1\,(head\, and\, ions) & -1.0 & 23.592 & 4.352\\
 & 1.0 & 27.057 & 4.352\\
slab\,2\,(head\, and\, ions) & 1.0 & -27.057 & 4.352\\
 & -1.0 & -23.592 & 4.352\end{array}$\\

In this bilayer model the potential field and forces at the core (Fig.
\ref{cap:Prof.fz-400chi} (a) and (b)) are higher than those of the
preceding section, due to the large values of the charge density $\rho(z)$
and the lack of the electrostatic shield provided by the water layers. 

\begin{figure}
\includegraphics[scale=0.3]{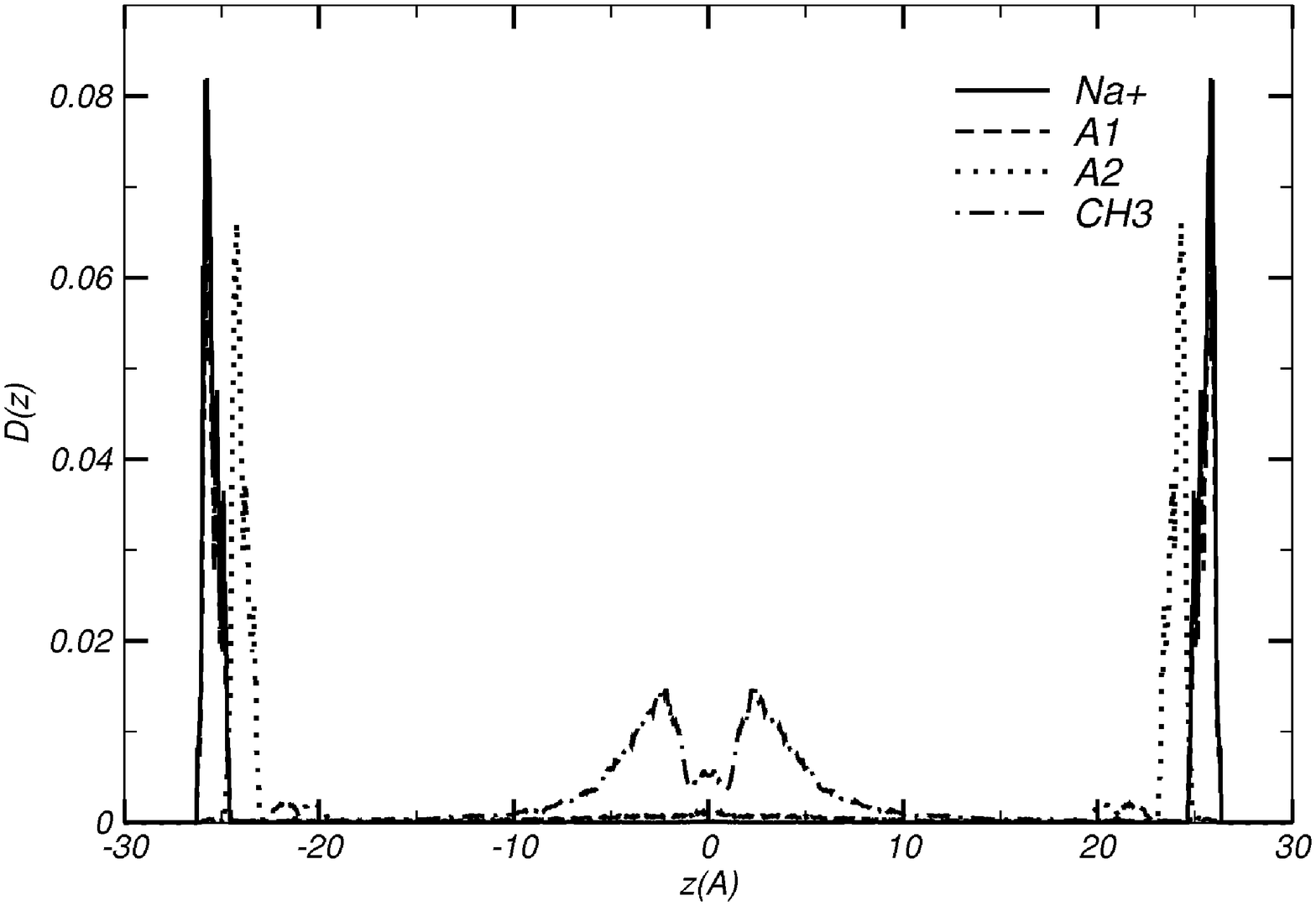}$\quad$\includegraphics[scale=0.3]{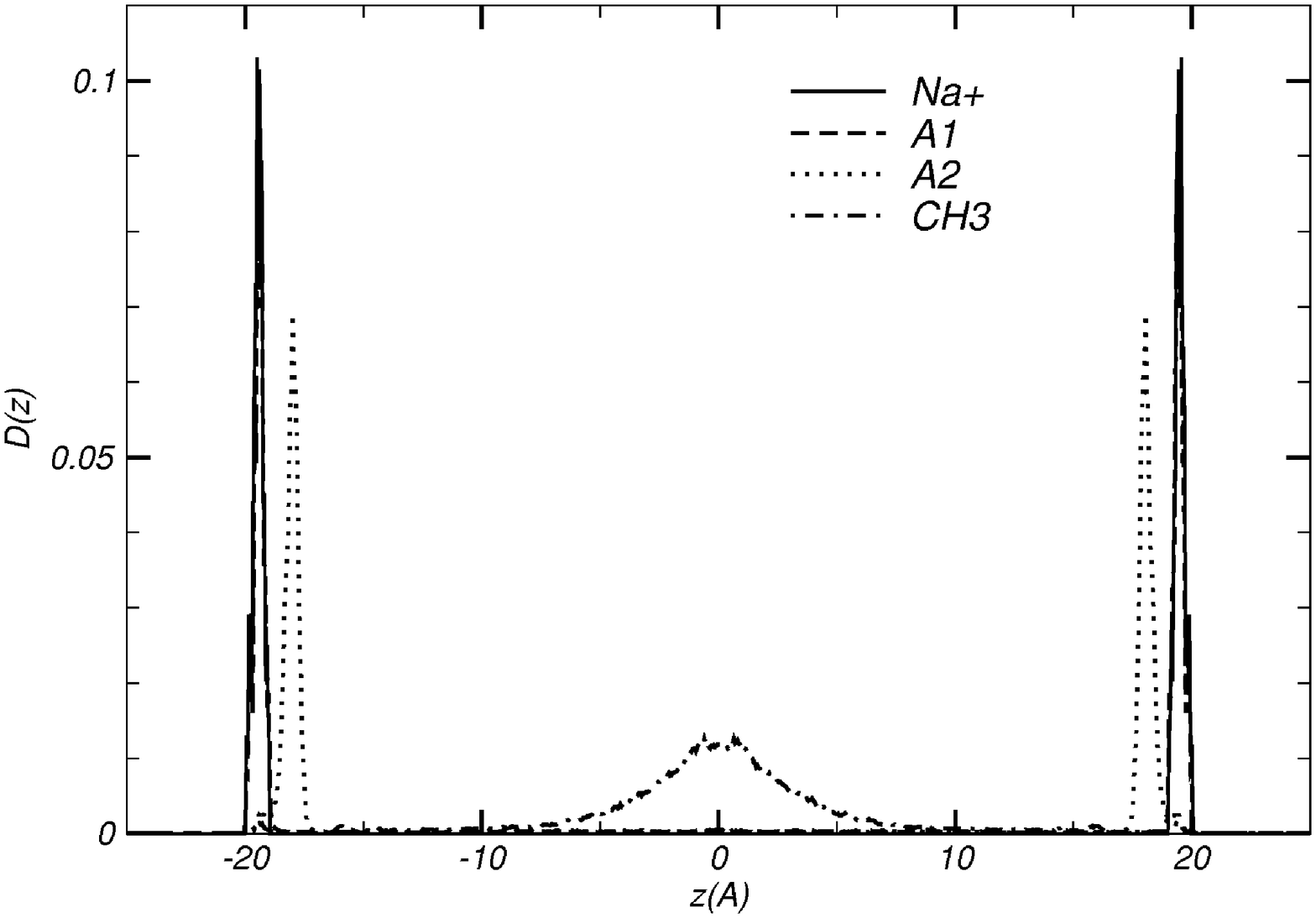}

\caption{Atomic density profiles (number of atoms per $\textrm{Å}³$), (a)
HD sample , (b) LD sample.\label{cap:mems-400chi-at.dens.prof} \protect \\
}
\end{figure}

The atomic density profiles (Fig. \ref{cap:mems-400chi-at.dens.prof})
show that the monoatomic positive ions remain at the border of the
slabs, and the same happens with most of the amphiphilic heads, although
for some of them we measured large excursions of the head groups within
the bilayers, as shown in Fig. \ref{cap:mem-400-conf}.

\begin{figure}
\vspace{2cm}\includegraphics[scale=0.3]{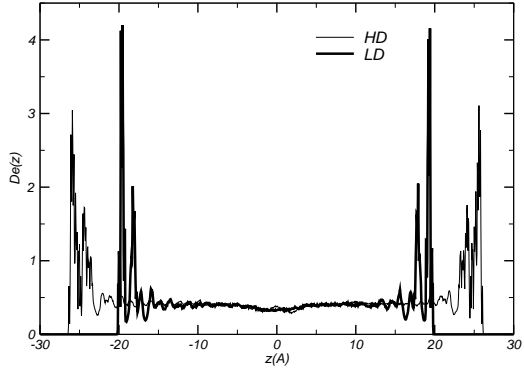}

\caption{Electron density profile (electrons/$\textrm{Å}³$) for HD and LD
head groups.\label{cap:mem-400chi-e.des.prof}\protect \\
}
\end{figure}

In the electron density profiles (Fig. \ref{cap:mem-400chi-e.des.prof})
we can see that the heavier atoms remain, on average, at less than
$\sim5\textrm{Å}$ of up and lower border.

\begin{figure}
[!ht]
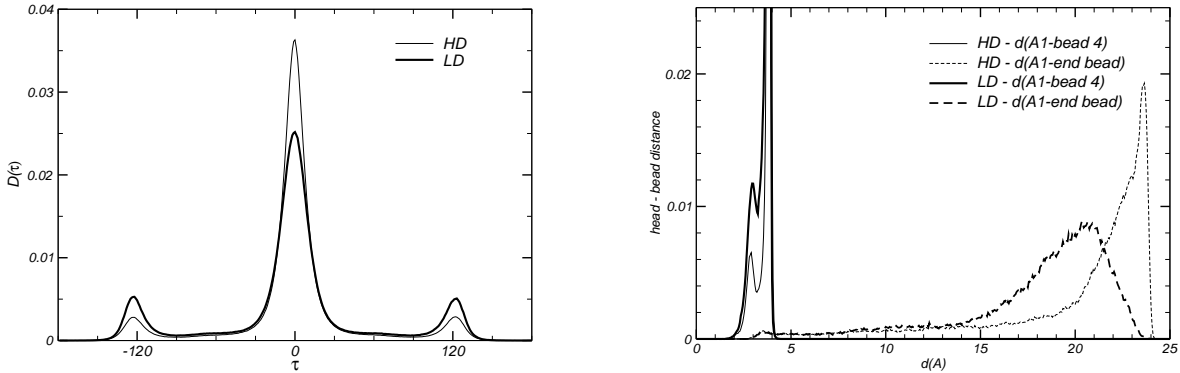
\vspace{1cm}\includegraphics[scale=0.3]{mems-chi-sinw-400c20-a56\lyxdot y\lyxdot a65-torsion}$\qquad\qquad$\includegraphics[scale=0.3]{mems-chi-sinw-400c20-a56\lyxdot y\lyxdot a65-fdc}

\caption{(a) Torsion angle distribution in the HD and LD samples. (b) The
head to bead 4 and head to tail average distances for amphiphilic
chains in HD and LD samples. \label{cap:Torsion-ang}\protect \\
}
\end{figure}

The distortion of the amphiphilic molecules is measured \emph{via}
the bending and torsional angles distribution (Fig. \ref{cap:Torsion-ang}(a))
and head to bead chain distances (Fig. \ref{cap:Torsion-ang}(b)).
From Fig. \ref{cap:Torsion-ang}(a) we estimate a 0.84 \emph{trans}
fraction of the torsional angles in the HD sample and a 0.74 fraction
in the LD sample. Accordingly, the head to bead 4 and head to tail
distances (Fig. \ref{cap:Torsion-ang}(b)) have a larger spread and
smaller chain length in the LD case.

\subsection{Simple biological membrane with surrounding water and neutral amphiphilics:}

\begin{figure}
[!ht]\includegraphics[scale=0.3]{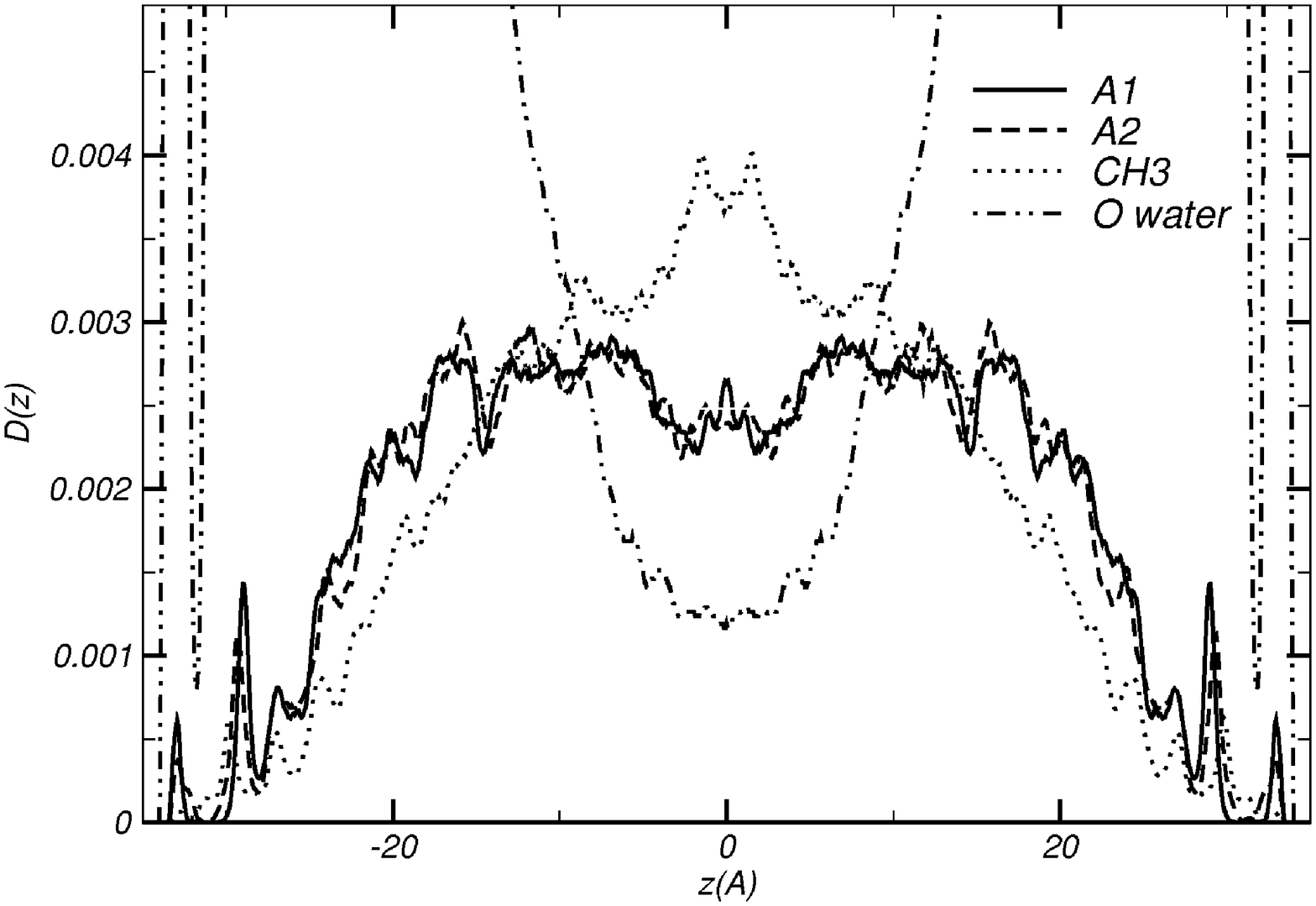}

\caption{Atomic density profile (number of atoms per $\textrm{Å}³$) for the
bilayer of neutral amphiphilics, without including the extra soft
hydrocarbon - hydrocarbon potential interaction term.\label{cap:At-dens-mems-cchwdip-sinvextrach23c}}
\end{figure}

The model membrane of this section consists of 226 neutral amphiphilics
and 2188 water molecules, the bilayer is perpendicular to the $\widehat{z}$
MD box axis, with a MD box size of $a=b=42.0$$\textrm{Å}$, $c=$1000
$\textrm{Å}$. The neutral amphiphilic is that described in section
III, a chain of 14 atoms where two of them model the dipolar head
(atom 1 with $q_{1}$ =- 1 e, atom 2 with $q_{2}$ = 1 e), sites 3
to 14 form the hydrophobic tail. The LJ atom - atom parameters are
equal in all of our amphiphilic models. In a constant volume MD simulation
we found that we cannot stabilize a bilayer structure at 300K, the
final sample is disordered, the water mix with the amphiphilics and
we calculate an almost identical distribution of heads and end groups
within the bilayer (Fig. \ref{cap:At-dens-mems-cchwdip-sinvextrach23c}).
This atomic profile, the measured diffusion coeficients and molecular
distortions (a trans fraction of 0.66\%) indicates a large disorder
in the calculated sample at 300K. \\

We conclude that with this neutral amphiphilic model the electrostatic
interactions are not strong enough to ensure a final bilayer structure
at 300K. This problem was already found in other simulations of simple
model bilayers without electrostatic interactions, as discused, for
example, in Ref. \cite{mem-potextra1}. 

The usual solution is to add a very soft and long ranged attractive
interaction term between nonbonded hydrophobic tails \cite{farago,mem-marcus-coarse.grain,mem-coarse-lipowsky,mem-potextra1}.
In coarse grained membrane models without electrostatic interactions,
for example, this extra interaction term between nonbonded hydrophobic
beads, is crucial for stabilizing fluid bilayers. This extra potential
term is usually modelled with a very soft potential of the type LJ
2-1 type instead of the usual LJ 12-6 \cite{farago,mem-marcus-coarse.grain},
or by extending the range of the 12-6 LJ potential with a flat section
at the minimum \cite{mem-potextra1}, all of them show a broad atractive
minimum. The obtained results suggest that this term is capable of
forcing the lipid chains into gel-like conformations and tends to
order the amphiphilic tails, decrease the area per head group and
reduce their lateral diffusion coefficient. \\

For our cases of weak electrostatic interactions we decided to test
the contribution of such term. Our proposed soft potential for the
nonbonded united atom $CH_{2}-CH_{2},\, CH_{2}-CH_{3},\, CH_{3}-CH_{3}$
interactions is:

$\begin{array}{ll}
V_{LJ}(r) & =4\varepsilon[(\frac{\sigma}{r})^{12}-(\frac{\sigma}{r})^{6}]\\
V_{soft}(r) & =2\varepsilon[(\frac{\sigma*0.9}{r})^{12}+Erfc\\
V_{final}(r) & =\frac{1}{2}[V_{LJ}(r)+V_{soft}(r)]\end{array}(\frac{r}{\sigma}-1)$\\

The idea after this proposed $V_{soft}(r)$ function can be better
seen when recalling that

$\frac{\partial}{\partial r}Erfc(\frac{r}{\sigma}-1)=-\frac{2}{\sigma\sqrt{\pi}}\exp(-(\frac{r}{\sigma}-1)^{2})$,\\
which means a wide gaussian spread of the atractive interaction forces
around $\sigma$ distances. In radial coordinates, the final interaction
forces between non-bonded hydrophobic sites become:

$\begin{array}{ll}
F_{LJ}(r) & =\frac{24\varepsilon}{r}[2(\frac{\sigma}{r})^{12}-(\frac{\sigma}{r})^{6}]\\
F_{soft}(r) & =4\varepsilon[\frac{6}{r}(\frac{\sigma*0.9}{r})^{12}-\frac{1}{\sigma\sqrt{\pi}}\exp(-(\frac{r}{\sigma}-1)^{2})\\
F_{final}(r) & =\frac{1}{2}[F_{LJ}(r)+F_{soft}(r)]\end{array}$]\\

\begin{figure}
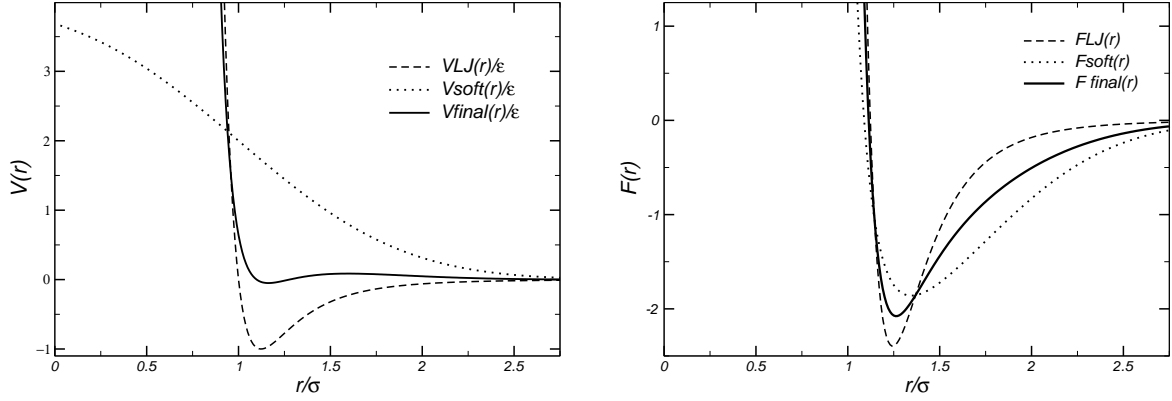

\includegraphics[scale=0.32]{pot\lyxdot v\lyxdot fextrach23c-plot}$\qquad$\includegraphics[scale=0.32]{pot\lyxdot f\lyxdot fextrach23c-plot}

\caption{Proposed soft potential term (a) and forces (b) between non bonded
united atoms $CH_{n}$\label{cap:soft-pot}.}
\end{figure}

Fig. \ref{cap:soft-pot} (a) includes the 12-6 LJ potential, the additional
soft term and the final potential model, Fig. \ref{cap:soft-pot}
(b) includes, in spherical coordinates, the corresponding forces.

\begin{figure}
\includegraphics[scale=0.3]{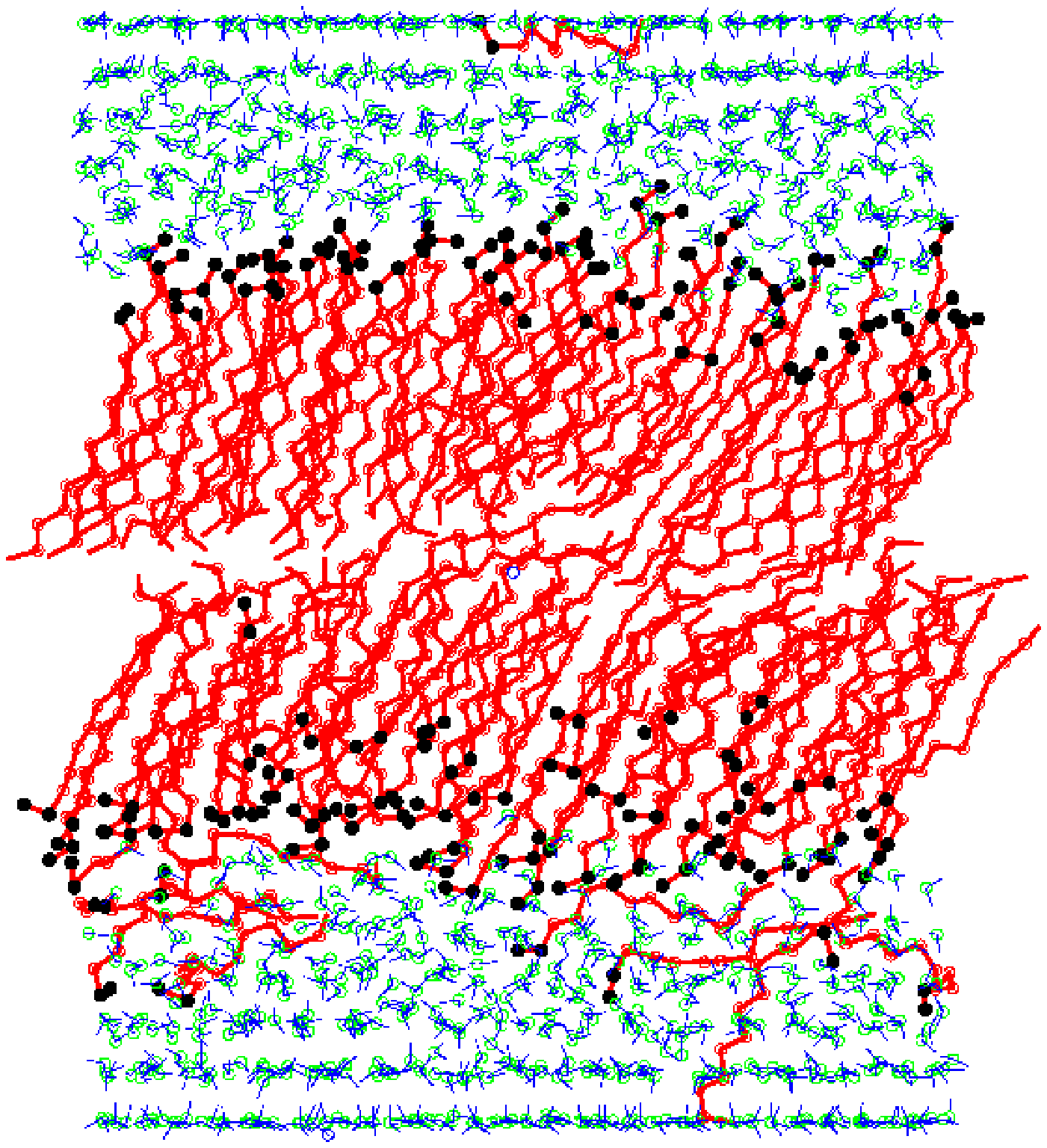}$\qquad\qquad$\includegraphics[scale=0.3]{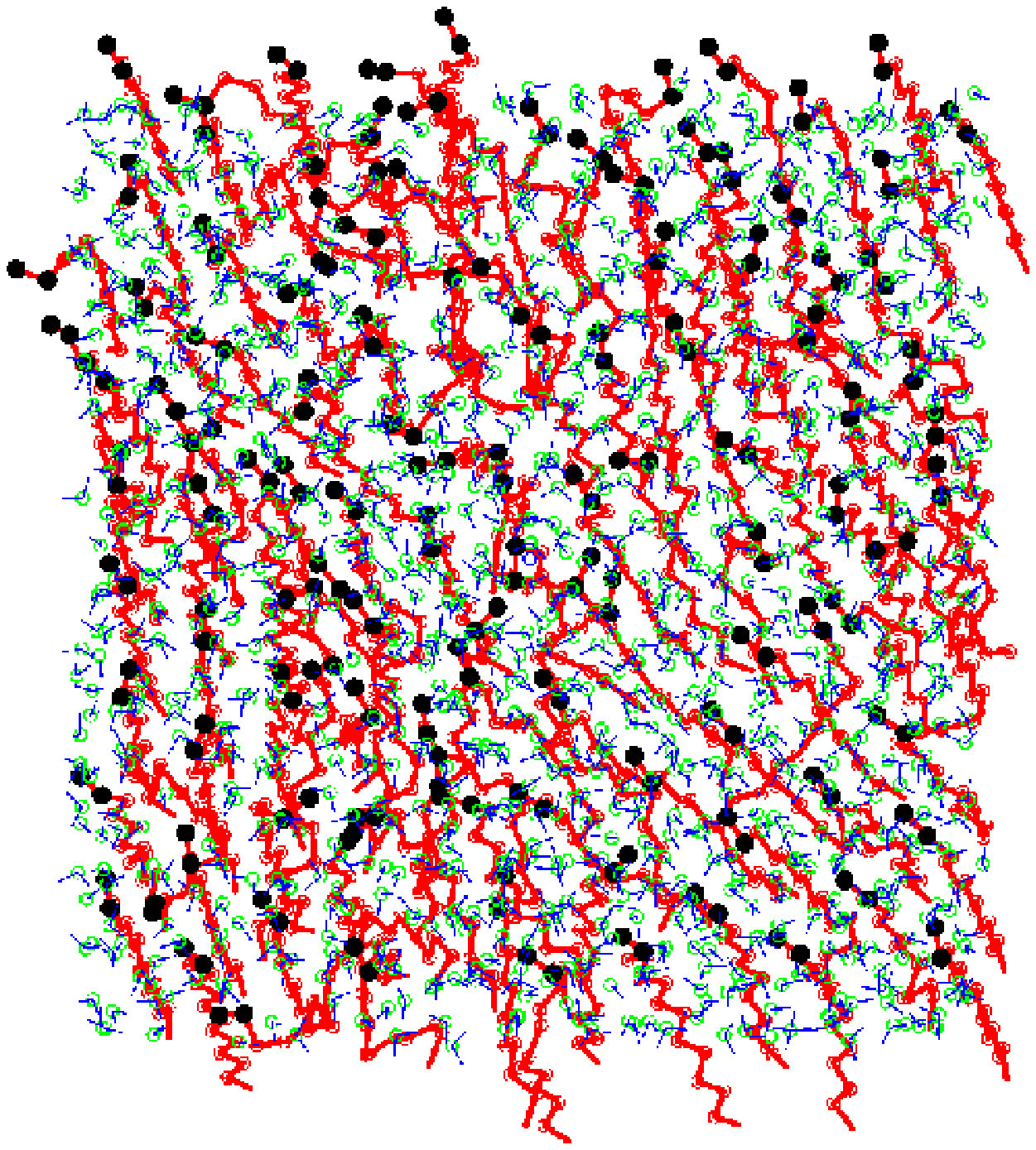}

\caption{Final configuration for the bilayer of neutral amphiphilics, including
our extra soft hydrocarbon- hydrocarbon potential interaction term.\label{capmems-:Final-conf-chwdip}}
\end{figure}

A new MD run on the same sample of neutral amphiphilics, but with
the addition of this soft potential term between non bonded united
atoms, was performed. Fig. \ref{capmems-:Final-conf-chwdip} shows
the equilibrated configuration. The calculated diffusion coefficient
is less than $10^{-7}cm^{2}/sec$, indicative of a gel phase. The
molecular distortions, as given by the \emph{trans} fraction of torsional
angles is 0.85\%.

\vspace{1cm}

\begin{figure}
\includegraphics[scale=0.3]{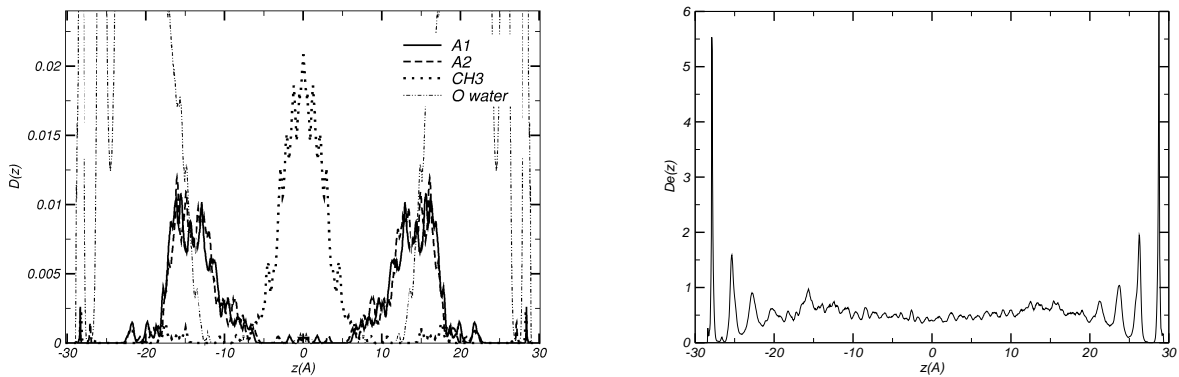}$\qquad\qquad$\includegraphics[scale=0.3]{mems-226c14-chwdip-a42-vextrach23c-a5-zdensel-angstrom3}

\caption{Atomic density profile (number of atoms per $\textrm{Å}³$) and electron
density profile (electrons/$\textrm{Å}³$) for the bilayer of neutral
amphiphilics, including our extra soft hydrocarbon- hydrocarbon potential
interaction term.\label{cap:At.dens.prof-chwdip}}
\end{figure}

\begin{figure}
\vspace{1cm}\includegraphics[scale=0.4]{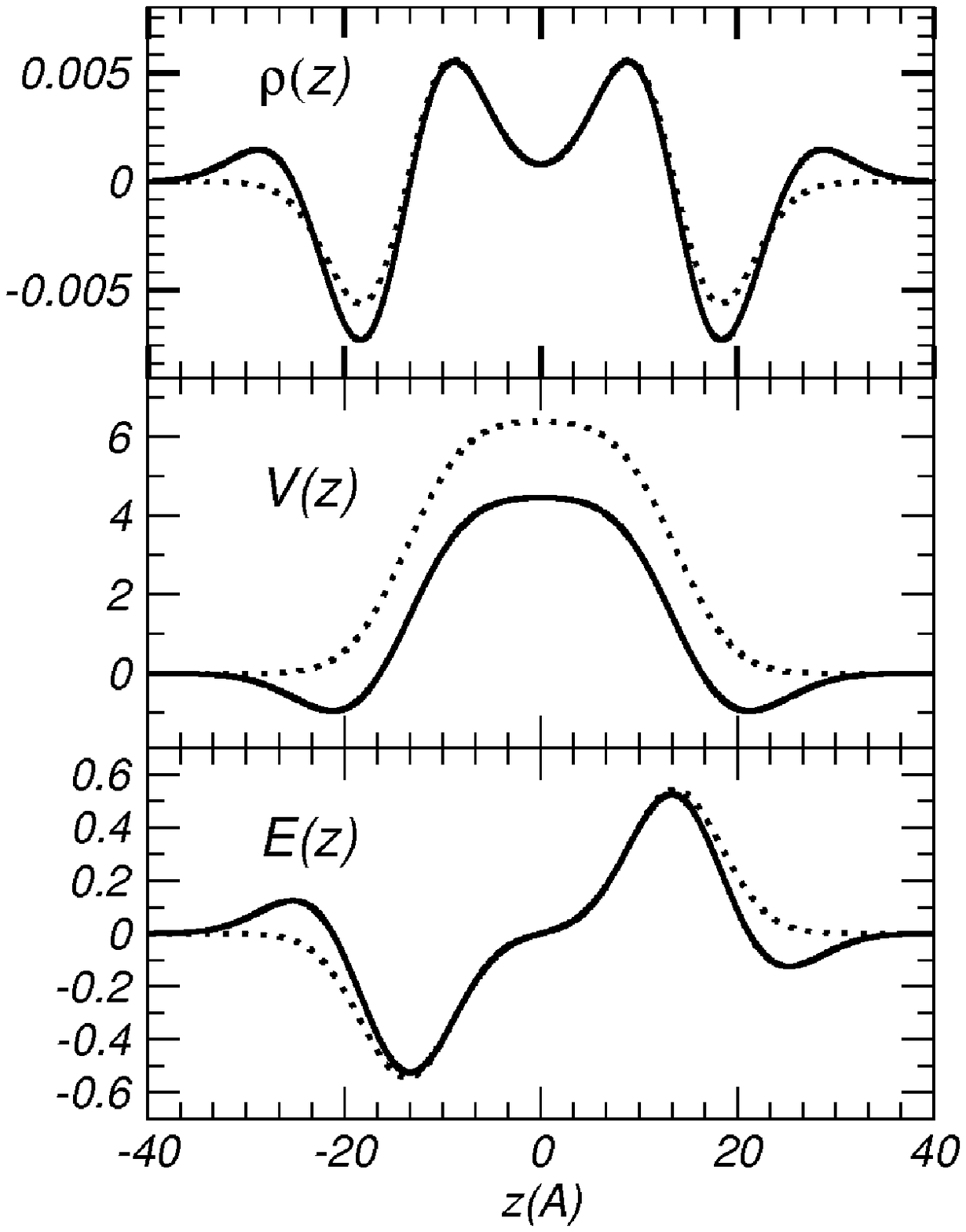}$\qquad\qquad\qquad\qquad\qquad$\includegraphics[scale=0.4]{mems-226c14-chwdip-a42-vextrach23c-a5-fzmido-chain}

\caption{Profile of: (a) the macroscopic charge density $\rho$(z), potential
$V(z)$ and electric field $E(z)$ (full line for total values, dotted
line for heads plus ions contribution), (b) our measured profile of
macroscopic electric, external and total forces on amphiphilics, as
a function of the location of their centers of mass along $z$, in
the model membrane of neutral amphiphilics, and including our extra
soft hydrocarbon- hydrocarbon potential interaction term.\label{cap:emacro-chwdip}\protect \\
}
\end{figure}

Fig. \ref{cap:At.dens.prof-chwdip} includes the atomic and electronic
density profile of this sample, and Fig. \ref{cap:emacro-chwdip}
its macroscopic electric field and forces on molecular centers of
mass, as a function of $\hat{z}$, the axis perpendicular to the bilayer.
Units as in the preceding sections. The macroscopic electric potential
\emph{V}(\emph{z}) is positive in the centre of the bilayer, because,
on time and spacial average, the head dipoles point to the interior
of the bilayer. \\

\begin{figure}
[!ht]\vspace{1cm}\includegraphics[scale=0.4]{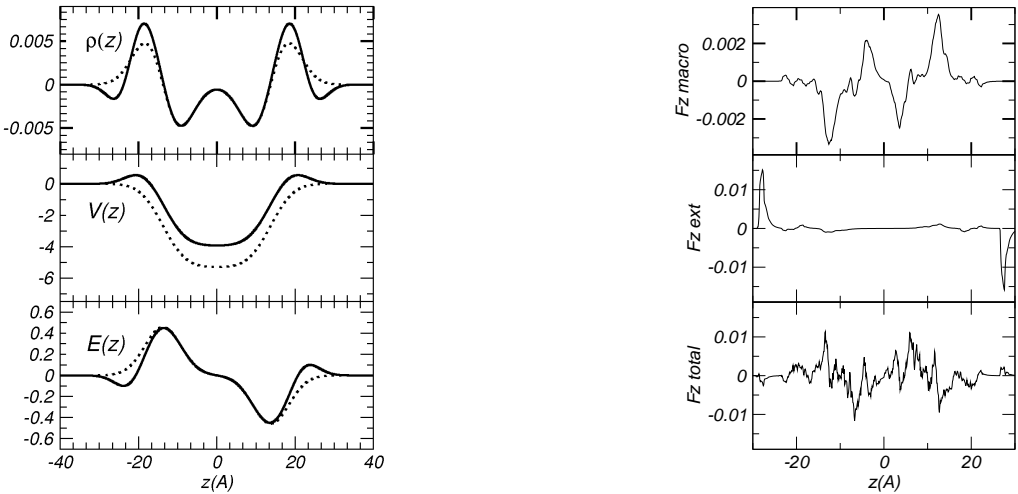}$\qquad\qquad\qquad\qquad\qquad$\includegraphics[scale=0.4]{mems-226c14-chwdip\lyxdot reves-a42\lyxdot 5-vextrach23c-e11-fzmido-chain-prom}

\caption{Profile of: (a) the macroscopic charge density $\rho$(z), potential
$V(z)$ and electric field $E(z)$ (full line for total values, dotted
line for heads plus ions contribution), (b) our measured profile of
macroscopic electric, external and total forces on amphiphilics, as
a function of the location of their centers of mass along $z$, in
the model membrane of neutral amphiphilics with reverse dipole (\emph{q(}A1\emph{)=1
e, q(}A2\emph{)=-1 e}), and including our extra soft hydrocarbon-
hydrocarbon potential interaction term.\label{cap:Prof-mem-neutra-a1+a2-}}
\end{figure}

This last point suggested a new MD run with model neutral amphiphilics
but with a reverse dipolar moment. Therefore the last studied model
membrane is entirely similar to the preceding one, except that the
neutral amphiphilics have a reverse dipolar head, that is, the A1
site has a charge of +1\emph{e} and site A2 has a charge of -1\emph{e}.
The soft potential term between non bonded united atoms is also included.
The procedure and the obtained results of this case are entirely similar
to those reported in this section, and we are not including them here,
except for the obtained profile of the macroscopic field and the measured
profile of forces on the amphiphilics as a function of the location
of their centers of mass along $z$, Fig.\ref{cap:Prof-mem-neutra-a1+a2-}.
The calculated macroscopic potential \emph{V}(\emph{z}) to that calculated,
for example, in the all-atom simulation of a membrane of neutral SDPC
lipids \cite{mem-electr-mike-1}, except that their negative potential
is of about -14.4$e/\textrm{Å}$, and ours is about -4$e/\textrm{Å}$.
The difference is explained by the average dipolar moment of the SDPC
molecule along \emph{z}, of about 0.9 $e\textrm{Å}$ and ours is about
0.2$e\textrm{Å}$ (due to a mean orientation angle of 82deg. with
respect to the bilayer normal).\\

The results of the last simulation implies that by increasing the
head dipolar moment of the neutral amphiphilics, the bilayer should
be stable. And effectively, with a MD simulation of neutral amphiphilics,
but with a strong charge of +\emph{4e} on site A1 and a charge of
-\emph{4e} on site A2, we obtained an equilibrated bilayer sample,
without the need of including the extra soft interaction term between
the non-bonded hydrocarbons sites.\\
 Most probably this is an indication that our membrane sample is not
fully hydrated and a larger amount of water should be included.

\section{Conclusion:}

In this paper we studied several simple models of amphiphilic biological
bilayers, and analyzed the key \emph{rôle} of the electrostatic interactions
in their self-assembly. The reverse model bilayer, a NB film was analysed
in a previous paper \cite{zg-bub08}. The molecular models are simple
enough so a large number of components can be included in the MD samples,
but also detailed enough so as to take into account molecular charge
distributions, flexible amphiphilic molecules and a reliable model
of water. All these properties are essential to obtain a reliable
conclusion at the nano- scale. Our amphiphilic model also allows to
study, in a simple way, the properties of bilayers formed by charged
or neutral amphiphilics and with or without explicity including water
molecules in the numerical simulations.\\

As for the calculation of the electrostatic interactions, we use our
proposed novel and more accurate method to calculate the macroscopic
electric field in cuasi 2D geometries \cite{zg-bub08}, which can
be easily included in any numerical calculation. The method, that
essentially is a coarsed grain fit of the macroscopic electric field
beyond the dipole order approximation, was applied here to symmetrical
bilayers (along the normal to the bilayer and with periodic boundary
conditions in two dimensions), but their derivation is general and
valid also for asymmetrical slab geometries.\\

We also propose a mean field method to take into account the far distant
water molecules interacting with a single bilayer of the biological
type. This procedure allows the study of one isolated biological bilayer
in solution, and not the usual stack of bilayers, as obtained when
3D periodic boundary conditions are applied. \\

Lastly, we emphasize the relevance and utility of these simple bilayer
models. They can be applied to the systematic study of the physical
properties of these bilayers, that strongly depend not only on 'external
parameters', like surface tension and temperature, but on the kind
of guest molecules of relevance in biological and/or ambient problems
embedded in them. In turn, the structure and dynamics of the embedded
molecules strongly depend on their interactions with the bilayer and
with the sorrounding water. 

These simple model bilayers can also be useful to model, for example,
the synthesis of inorganic (ordered or disordered) materials \emph{via}
an organic agent \cite{nanotech}. This is a recent and very fast
growing research field of nanotechnologycal relevance, as are lithography,
etching and molding devices at the nanoscopic scale. Another fast
developing field is that of electronic sensors and nanodevices supported
on lipid bilayers \cite{nanotech2,nanotech3}. Lastly, as our model
retains the flexibility of the original amphiphilics, and the electrostatic
interactions are included, the approach is really useful to obtain
'realistic' solutions to the above mentioned problems as well as those
related to electric fields and electrostatic properties.\\

\begin{acknowledgments}
Z. G. greatly thanks careful reading and helpful sugestions to J.
Hernando.
\end{acknowledgments}

\end{document}